\documentclass[a4paper,11pt]{amsart}
\usepackage{times,epsfig,amssymb}

\renewcommand{\theequation}{\arabic{section}.\arabic{equation}}

\newcommand{\Z}{\mathbb{Z}}
\newcommand{\R}{\mathbb{R}}

\newcommand{\N}{\mathbb{N}}
\renewcommand{\L}{\mathbb{L}}
\renewcommand{\P}{\mathbb{P}}
\newcommand{\E}{\mathbb{E}}

\newcommand{\cL}{\mathcal{L}}

\newcommand{\cJ}{\mathcal{J}}
\newcommand{\cF}{\mathcal{F}}
\newcommand{\cQ}{\mathcal{Q}}

\newcommand{\Int}{\mathrm{Int}}

\newcommand{\meas}{\mathrm{meas}}
\newcommand{\supp}{{\ensuremath{\rm supp}}}
\newcommand{\dist}{{\ensuremath{\rm dist}}}
\newcommand{\diam}{{\ensuremath{\rm diam}}}

\DeclareMathOperator*{\ulim}{\overline{lim}}
\DeclareMathOperator*{\llim}{\underline{lim}}
\DeclareMathOperator{\tr}{tr}
\DeclareMathOperator{\spec}{spec}

\newtheorem{theorem}{Theorem}[section]{\bf}{\it}
\newtheorem{proposition}[theorem]{Proposition}{\bf}{\it}
\newtheorem{corollary}[theorem]{Corollary}{\bf}{\it}
\newtheorem{example}[theorem]{Example}{\it}{\rm}
\newtheorem{lemma}[theorem]{Lemma}{\bf}{\it}
\newtheorem{remark}[theorem]{Remark}{\it}{\rm}
\newtheorem{definition}[theorem]{Definition}{\bf}{\it}

\title{The Density of States and the Spectral Shift Density
of Random Schr\"{o}dinger Operators}

\author{V. Kostrykin \and R. Schrader}

\address{Vadim Kostrykin\\
Lehrstuhl f\"{u}r Lasertechnik\\ Rheinisch
- Westf\"{a}lische Technische Hoch\-schule Aachen\\
Steinbachstr. 15, D-52074 Aachen, Germany}
\email{kostrykin@t-online.de, kostrykin@ilt.fhg.de}

\address{Robert Schrader\\ Institut f\"{u}r
Theoretische Physik\\ Freie Universit\"{a}t Berlin, Arnimallee 14\\ D-14195 Berlin,
Germany}
\email{schrader@physik.fu-berlin.de}

\thanks{$^\ast$ R.S. supported in part by
DFG SFB 288 ``Differentialgeometrie und Quantenphysik''}
\thanks{To appear in
\textit{Reviews in Mathematical Physics}}



\date{February 10, 1999}

\begin{document}

\begin{abstract}
In this article we continue our analysis of Schr\"{o}dinger operators with a random
potential using scattering theory. In particular the theory of Krein's spectral
shift function leads to an alternative construction of the density of states in
arbitrary dimensions. For arbitrary dimension we show existence of the spectral
shift density, which is defined as the bulk limit of the spectral shift
function per unit interaction volume. This density equals the difference of the
density of states for the free and the interaction theory. This extends the
results previously obtained by the authors in one dimension. Also we consider
the case where the interaction is concentrated near a hyperplane.
\end{abstract}

\maketitle
\thispagestyle{empty}

\section{Introduction}

The integrated density of states is a quantity of primary interest in the
theory and in applications of one-particle random Schr\"{o}dinger operators. In
particular the topological support of the associated measure coincides with the
almost-sure spectrum of the operator. Moreover, its knowledge allows to compute
the free energy and hence all basic thermodynamic quantities of the
corresponding non-interacting many-particle systems.

The present article is a continuation of our analysis of applications of
scattering theory to random Schr\"{o}dinger operators
\cite{Kostrykin:Schrader:98a,Kostrykin:Schrader:98b}. There we showed in
particular in the one-dimensional context the existence of the bulk limit of
the spectral shift function per unit interaction interval. Also this limit was
shown to be equal to the difference of the integrated densities of states for
the free and the interaction theory. Here we extend this result to arbitrary
dimensions $\nu$. This result was announced in \cite{Kostrykin:Schrader:98a}.
An independent proof has been recently given in \cite{Chahrour} in the case of
the discrete Laplacian. In \cite{Kostrykin:Schrader:98a} we also proved how the
Lyapunov exponent could be obtained in an analogous way as (minus) the bulk
limit for the logarithm of the absolute value of the scattering amplitude per
unit interaction interval. This result was recognized long ago, although a
complete proof was absent, see
\cite{Lifshitz:Gredeskul:Pastur:82,Lifshitz:Gredeskul:Pastur:88}. We believe
that a similar result can be obtained for the higher dimensional case (see
\cite{Kostrykin:Schrader:98a} for a precise formulation).

Some other applications of scattering theory in one dimension to the study of
spectral properties of Schr\"odinger operators with periodic or random
potentials can be found in \cite{Keller,Rorres,Pavlov:Smirnov} and
\cite{Kirsch:Kotani:Simon} respectively.

One of the important ingredients of our approach is the Lifshitz-Krein spectral
shift function (see \cite{Birman:Yafaev,Birman:Pushnitski} for a review and
\cite{Gesztesy:Makarov:Naboko,Gesztesy:Makarov} for recent results). In the
context of our approach the spectral shift function naturally replaces the
eigenvalue counting function, which is usually used to construct the density of
states. The celebrated Birman-Krein theorem \cite{Birman:Krein} relates the
spectral shift function to scattering theory. In fact, up to a factor
$-\pi^{-1}$ it may be identified with the scattering phase when the energy
$\lambda>0$. For $\lambda<0$ the spectral shift function equals minus the
eigenvalue counting function.

These two properties of the spectral shift function, namely its relation to
scattering theory and its replacement of the counting function in the presence
of an absolutely continuous spectrum convinced the authors already some time
ago that the spectral shift function could be applied to the theory of random
Schr\"odinger operators and led us to an investigation of cluster properties of
the spectral shift function \cite{Kostrykin:Schrader:94,Kostrykin:Schrader:98},
when the potential is a sum of two terms and the center of one is moved to
infinity. In \cite{Geisler:Kostrykin:Schrader} we proved convexity and
subadditivity properties of the integrated spectral shift function with respect
to the potential and the coupling constant, respectively. Such properties often
show up when considering thermodynamic limits in statistical mechanics.

In the one-dimensional case \cite{Kostrykin:Schrader:98a} we proved an
inequality for the spectral shift function, which reflect its ``additivity''
properties with respect to the potential being the sum of two terms with
disjoint supports
\begin{displaymath}
|\xi(\lambda;H_0+V_1+V_2,H_0)-\xi(\lambda;H_0+V_1,H_0)-\xi(\lambda;H_0+V_2,H_0)|\leq
1.
\end{displaymath}
Combined with the superadditive (Akcoglu--Krengel) ergodic theorem
\cite{Krengel} this allowed us to prove for random Hamiltonians of the form
\begin{displaymath}
H_\omega^{(n)}=H_0+ \sum_{j=-n}^{j=n}\alpha_j(\omega) f(\cdot-j)
\end{displaymath}
the almost sure existence of the limit
\begin{equation}\label{xi}
\xi(\lambda)=\lim_{n\rightarrow\infty}\frac{\xi(\lambda;H_\omega^{(n)},H_0)}{2n+1},
\end{equation}
which we called the spectral shift density. We proved the equality
$\xi(E)=N_0(E)-N(E)$, where $N(E)$ and $N_0(E)=\pi^{-1}[\max(0,E)]^{1/2}$ are
the integrated density of states of the Hamiltonians $H(\omega)$ and $H_0$
respectively.

Before we outline the main results of this paper we recall some well-known
facts about the density of states for Schr\"{o}dinger operators $H=H_0+V$ in the
Hilbert space $L^2(\R^{\nu})$ with $H_0=-\Delta$ and $V$ being an arbitrary
potential with $V_-\in K_{\nu}$, $V_+\in K^{\mathrm{loc}}_{\nu}$ ($K_\nu$
denotes here the Kato class, see e.g.
\cite{Cycon:Froese:Kirsch:Simon,Simon:82}). One says that $H=H_0+V$ has a
\emph{density of states measure} if for all $g\in C_0^\infty(\R)$
\begin{equation}\label{DOS.1}
\mu(g)=\lim_{\Lambda\rightarrow\infty}\tr(\chi_\Lambda g(H))/\meas(\Lambda)
\end{equation}
exists. Here $\chi_\Lambda$ is the characteristic function of a rectangular box
$\Lambda=[a_1,b_1]\times\ldots\times [a_\nu,b_\nu]$ and the limit
$\Lambda\rightarrow\infty$ is understood in the sense $a_i\rightarrow -\infty$,
$b_i\rightarrow\infty$ for all $i=1,\ldots,\nu$. Actually $\Lambda$ need not be
a box. Instead of boxes we can take a sequence $\Lambda_i$ of bounded domains
tending to infinity in the sense of Fisher \cite{Pastur:Figotin}. With
$\Lambda^{(h)}$ being the set of points within distance $h$ from the boundary
$\partial\Lambda$ of $\Lambda$, the convergence in the sense of Fisher means
that $\lim\meas(\Lambda_i)=\infty$ and for any $\epsilon>0$ there exists
$\delta>0$ independent of $i$ and such that $\meas(\Lambda_i^{(\delta\
 \diam(\Lambda_i))})/\meas(\Lambda_i)<\epsilon$.

By the Stone-Weierstrass theorem for the existence of the density of states
measure it suffices to prove the existence of the limit on the r.h.s.\ of
\eqref{DOS.1} with $g(\lambda)=e^{-\lambda t}$ for all $t>0$ \cite{Simon:82}.

By Riesz's representation theorem the positive linear functional $\mu(g)$
defines a positive Borel measure $dN(E)$ (density of states measure) such that
\begin{equation}\label{DOS.2}
\mu(g)=\int_\R g(\lambda)dN(\lambda).
\end{equation}
The non-decreasing function
\begin{displaymath}
N(\lambda)=\int_{-\infty}^{\lambda-0}dN(\lambda')\equiv N((-\infty,\lambda))
\end{displaymath}
is called the \emph{integrated density of states}. If the density of states
measure is absolutely continuous, its Radon-Nikodym derivative $n(E)=dN(E)/dE$
is called the \emph{density of states}. For random Schr\"{o}dinger operators the
absolute continuity of $N(E)$ is discussed in
\cite{Kotani:Simon:87,Combes:Hislop:94,Barbaroux:Combes:Hislop:97,Fischer:Hupfer:Leschke:Mueller}.

Let $H_\Lambda^{D}$ be the operator $H_{0,\Lambda}^D+V$ where $H_{0,\Lambda}^D$
is the Laplacian on $L^2(\Lambda)$ with Dirichlet boundary conditions on
$\partial
\Lambda$. Then
\begin{equation}\label{DOS.4}
\lim_{\Lambda\rightarrow\infty}(\meas(\Lambda))^{-1}
\left[\tr(\chi_\Lambda g(H))-\tr(g(H_\Lambda^D)) \right]=0,
\end{equation}
such that the integrated density of states can be calculated as the bulk limit
of the density of the eigenvalue counting function for $H_\Lambda^{D}$. This
equation shows that the limit (\ref{DOS.1}) does not depend on the properties
of $H$ ``outside" the box $\Lambda$. Therefore one may expect that
\begin{equation}\label{DOS.5}
\lim_{\Lambda\rightarrow\infty}(\meas(\Lambda))^{-1}
\left[\tr(\chi_\Lambda g(H))-\tr(\chi_\Lambda g(H_0+\chi_\Lambda V))
\right]=0.
\end{equation}
Below we will prove (see Theorem \ref{th.SSD.1}) that this really is the case.
Substracting from (\ref{DOS.1}) the same limit with $H=H_0$, i.e.\ $V=0$ und
using (\ref{DOS.5}) we obtain
\begin{equation}\label{DOS.6}
\mu(g)-\mu_0(g)=\lim_{\Lambda\rightarrow\infty}(\meas(\Lambda))^{-1}\left[
\tr(\chi_\Lambda g(H_0+\chi_\Lambda V))-\tr(\chi_\Lambda g(H_0))\right].
\end{equation}
By construction the potential $\chi_\Lambda V$ has compact support. This fact
will allow us to prove that the difference $g(H_0+\chi_\Lambda V)-g(H_0)$ is
trace class for all finite $\Lambda$. Since $g(H_0+\chi_\Lambda V)$ outside the
box $\Lambda$ is ``approximately" equal to $g(H_0)$ we will be able to prove
that
\begin{equation}\label{DOS.7}
\lim_{\Lambda\rightarrow\infty}(\meas(\Lambda))^{-1}\tr\left[(1-\chi_\Lambda)
(g(H_0+\chi_\Lambda V)-g(H_0)) \right]=0.
\end{equation}
Combining (\ref{DOS.6}) and (\ref{DOS.7}) we obtain
\begin{eqnarray}
\mu(g)-\mu_0(g)&=&\lim_{\Lambda\rightarrow\infty}(\meas(\Lambda))^{-1}
\tr\left[g(H_0+\chi_\Lambda V)-g(H_0) \right]\nonumber\\
&=&\lim_{\Lambda\rightarrow\infty}\int_{\R} g'(\lambda)
\frac{\xi(\lambda;H_0+\chi_\Lambda V,H_0)}{\meas(\Lambda)}d\lambda,\label{DOS.9}
\end{eqnarray}
where $\xi(\lambda; H_0+\chi_\Lambda V, H_0)$ is the spectral shift function
for the pair of operators ($H_0+\chi_\Lambda V$, $H_0$).

Since the l.h.s.\ of (\ref{DOS.9}) is a difference of two positive linear
functionals, the existence of the density of states implies the existence of a
limiting (signed) measure $d\Xi(\lambda)$ such that
\begin{displaymath}
\int_\R g(\lambda) d\Xi(\lambda) = \lim_{\Lambda\rightarrow\infty}\int g(\lambda)
\frac{\xi(\lambda;H_0+\chi_\Lambda V, H_0)}
{\meas(\Lambda)}d\lambda
\end{displaymath}
for any $g\in C_0^1$ (continuously differentiable functions with compact
support). Also, from (\ref{DOS.2}) and (\ref{DOS.9}) it follows that
\begin{equation}\label{DOS.10}
\int g(\lambda)dN(\lambda)-\int g(\lambda)dN_0(\lambda)=\int g'(\lambda)d\Xi(\lambda).
\end{equation}
Since $N(\lambda)$ and $N_0(\lambda)$ are both non-decreasing functions we may
view the integrals on the l.h.s.\ of (\ref{DOS.10}) as Lebesgue-Stieltjes
integrals and perform an integration by parts, thus obtaining
\begin{equation}\label{DOS.10.2}
\int g'(\lambda)(N_0(\lambda)-N(\lambda))d\lambda = \int g'(\lambda) d\Xi(\lambda).
\end{equation}
This implies that $d\Xi(\lambda)$ is absolutely continuous. Its Radon-Nikodym
derivative $\xi(\lambda)=d\Xi(\lambda)/d\lambda$ we call the \emph{spectral
shift density}. From (\ref{DOS.10.2}) we also have
\begin{equation}\label{DOS.11}
\xi(\lambda)=N_0(\lambda)-N(\lambda)\quad \textrm{a.e.\ on}\ \R.
\end{equation}
Clearly the converse is also true, i.e.\ if the spectral shift density exists
then the density of states also exists and (\ref{DOS.11}) is fulfilled.

Similarly we can prove the existence of the relative spectral shift density
\begin{displaymath}
\lim_{\Lambda\rightarrow\infty}\int g(\lambda)
\frac{\xi(\lambda;H_0+\chi_\Lambda V +W, H_0+W)}{\meas(\Lambda)}d\lambda,
\end{displaymath}
which is again related to the difference of the densities of states for the
operators $H_0+V+W$ and $H_0+W$. For example as in
\cite{Hempel:Kirsch,Barbaroux:Combes:Hislop:97} we can take $W$ to be a
periodic potential and $V$ to be a random potential describing the distribution
of impurities. We expect that it is also possible to consider Schr\"{o}dinger
operators with an electromagnetic field
\begin{displaymath}
H_0(a)= (-i\nabla+a)^2 +W,
\end{displaymath}
where $a$ is a vector potential of a magnetic field and $W$ stands for an
electrostatic potential. However, we will not touch this question in the
present work.

The heuristic consideration presented above will be rigorously justified in
Section 2. In Section 3 we will show that actually it is not necessary to take
a ``sharp" cut-off $\chi_\Lambda V$ to calculate the spectral shift density.
For lattice-type potentials of the form $V=\sum_{\mathbf{j}\in\Z^\nu}
f_{\mathbf{j}}(\cdot-\mathbf{j})$, where
$\{f_{\mathbf{j}}\}_{\mathbf{j}\in\Z^\nu}$ is a family of not necessarily
compactly supported functions being uniformly in the Birman-Solomyak class
$l^1(L^2)$, one can approximate $V$ by a sequence of
$V_\Lambda=\sum_{\mathbf{j}\in\Lambda}f_{\mathbf{j}}(\cdot-\mathbf{j})$.
Section 4 is devoted to the study of the cluster proprties for the Laplace
transform of the spectral shift function (see Corollary \ref{cor:extension}).

In Section 5 we consider random Schr\"{o}dinger operators of two types, namely the
random crystal model,
\begin{equation}\label{type1}
H_\omega= H_0 +\sum_{\mathbf{j}\in\Z^\nu}\alpha_{\mathbf{j}}(\omega)
f(\cdot-\mathbf{j}),
\end{equation}
and that of a monoatomic layer
\begin{equation}\label{type2}
H_\omega= H_0 +\sum_{\mathbf{j}\in\Z^{\nu_1}}\alpha_{\mathbf{j}}(\omega)
f(\cdot-\mathbf{j}),\quad \nu_1<\nu,
\end{equation}
where $f$ is supposed to be compactly supported on the unit cell and
$\alpha_{\mathbf{j}}(\omega)$ is a sequence of random i.i.d.\ variables forming
a stationary metrically transitive field. For the Hamiltonians (\ref{type1})
the existence of the integrated density of states $N(\lambda)$ is well known
(see e.g. \cite{Kirsch:Martinelli:82a}). We prove that for any $g\in C_0^1$
\begin{displaymath}
\lim_{\Lambda\rightarrow\infty} \int g(\lambda)\frac{\xi(\lambda;H_0+V_{\omega,\Lambda},H_0)}
{\meas(\Lambda)}d\lambda=\int g(\lambda)(N_0(\lambda)-N(\lambda))d\lambda
\end{displaymath}
almost surely. This result also remains valid for Hamiltonians of the form
$H_\omega=H_0+V_\omega$, where $V_\omega(x)$ is an arbitrary metrically
transitive random field, i.e.\ there are measure preserving ergodic
transformations $\{T_y\}_{y\in\R^\nu}$ such that
$V_{T_y\omega}(x)=V_\omega(x-y)$.

For the Hamiltonians of the type \eqref{type2} we prove the existence of the
spectral shift density as a measure (see Theorem \ref{SSDens:8} below).
Recently similar results for discrete Schr\"{o}dinger operators of this type were
obtained by A. Chahrour in \cite{Chahrour}.

\textbf{Acknowledgements.} We are indebted to A. Chahrour, J.M. Combes, V. Enss, W. Kirsch
and L. Pastur for useful discussions.

%%%%%%%%%%%%%%%%%%%%%%%%%%%%%%%%%%%%%%%%%%%%%%%%%%%%%%%%%%%%%%
\section{Spectral Shift Density: General Potentials}
\setcounter{equation}{0}
%%%%%%%%%%%%%%%%%%%%%%%%%%%%%%%%%%%%%%%%%%%%%%%%%%%%%%%%%%%%%%

We start with some preparations. Let $(\boldsymbol{\Omega}_x,\ \mathbf{P}_x,
(X_t)_{t\geq 0})$ denote the Brownian motion starting at $x\in\R^\nu$ with
expectation $\mathbf{E}_x$. For an arbitrary measurable set $B\subset\R^\nu$
let $\tau_B(\boldsymbol{\omega})$,
$\boldsymbol{\omega}\in\boldsymbol{\Omega}_x$ be the first hitting time:
\begin{displaymath}
\tau_B(\boldsymbol{\omega})=\inf_{t>0}\{X_t(\boldsymbol{\omega})\in B \}.
\end{displaymath}

Let $\cJ_1$ and $\cJ_2$ denote the ideals of trace class and Hilbert-Schmidt
operators in the Hilbert space $L^2=L^2(\R^\nu)$ with norms $\|\cdot\|_{\cJ_1}$
and $\|\cdot\|_{\cJ_2}$ respectively. Also for any potential $V$, $V_+$ and
$V_-$ are its positive and non-positive parts respectively such that $V=V_+
+V_-$. The following theorem was proven by Stollmann in \cite{Stollmann:94}
(see also \cite{Stollmann:93}, where these results were announced).

\begin{theorem}\label{thm:Stollmann:1}
Let $V,W$ be such that $V_+, W_+\in K_\nu^{\mathrm{loc}}$, $V_-, W_-\in K_\nu$
and $V$ has compact support. Then
\begin{eqnarray}
\left\|e^{-t(H_0+W+V)}-e^{-t(H_0+W)}\right\|_{\cJ_2}&\leq&
c_2\left\|\mathbf{P}_\bullet\{\tau_{\supp V}\leq t/2 \}
\right\|_{L^1}^{1/2},\label{St:gl:1}\\
\left\|e^{-t(H_0+W+V)}-e^{-t(H_0+W)}\right\|_{\cJ_1}&\leq&
c_1\left\|\mathbf{P}_\bullet\{\tau_{\supp V}\leq t/2 \}^{1/2}
\right\|_{L^1},\label{St:gl:2}
\end{eqnarray}
\end{theorem}

\begin{remark}\label{rem:Stollmann:1}
Inspecting the proofs in \cite{Stollmann:94} one can easily see that the
constants $c_2$ and $c_1$ in \eqref{St:gl:1} and \eqref{St:gl:2} respectively
can be chosen as follows
\begin{eqnarray*}
c_2 &=& 8 (2\pi
t)^{-\nu/4}\left\|e^{-t(H_0+W_-+V_-)/2}\right\|_{\infty,\infty},\\ c_1 &=&
2^{3-\nu/4} (\pi t)^{-\nu/2}
\left\|e^{-t(H_0+2W_-+2V_-)/2}\right\|^{1/2}_{\infty,\infty}\
\left\|e^{-t(H_0+4W_-+4V_-)/4}\right\|^{1/2}_{\infty,\infty}.
\end{eqnarray*}
\end{remark}

Actually in \cite{Stollmann:94} this theorem was proven under the much more
general conditions on the perturbations $V$ and $W$ by which they were allowed
to be measures. In the sequel we will not use the Hilbert-Schmidt estimates.
Nevertheless we have included them since from our point of view they provide an
interesting information on the convergence of semigroup differences.

The following lemma allows one to estimate the r.h.s.\ of (\ref{St:gl:1}) and
of (\ref{St:gl:2}) in terms of $\meas(\supp V)$:

\begin{lemma}\cite{Stollmann:Stolz:94}\label{thm:Stollmann:2}
For an arbitrary measurable set $B\subset\R^\nu$ and for any $x\notin B$ such
that $\dist(x,B)>0$
\begin{displaymath}
\mathbf{P}_x\{\tau_B\leq t \} \leq 2\nu \exp\left\{-\frac{\dist(x,B)^2}{4\nu t} \right\}.
\end{displaymath}
\end{lemma}

Thus the r.h.s.\ of (\ref{St:gl:1}) can be bounded by $(\meas(\supp V))^{1/2}$
and the r.h.s.\ of (\ref{St:gl:2}) by $\meas(\supp V)$.

Let $\|A\|_{p,q}$ denote the norm of the operator $A$ as a map from $L^p$ into
$L^q$, $1\leq p,q\leq\infty$. Using some ideas and methods from
\cite{Stollmann:94} we will prove

\begin{theorem}\label{thm:Stollmann:3}
Let $B\subset\R^\nu$ be a compact set. Let $V$ be a measurable function such
that $V_+\in K_\nu^{\mathrm{loc}}$ and $V_-\in K_\nu$. Then for any $t>0$ there
is a constant $c>0$ independent of $B$ such that
\begin{eqnarray}
\left\|\chi_B\left(e^{-t(H_0+V)}-e^{-t(H_0+\chi_B V)}\right) \right\|_{\cJ_2}
\leq
2^{1-\nu/2}(\pi t)^{-\nu/4}\left\|e^{-t(H_0+2V_-)/2}
\right\|_{\infty,\infty}\nonumber\\ \cdot\left(3\|\chi_B\mathbf{P}_\bullet\{\tau_{B^c}\leq
t/2\}\|_{L^1}^{1/2}+\|\mathbf{E}_\bullet\{\chi_B(X_t);\tau_{B^c}\leq t/2 \}
\|_{L^1}^{1/2}\right),\label{St:gl:3}\\[4mm]
\left\|\chi_B\left(e^{-t(H_0+V)}-e^{-t(H_0+\chi_B V)}\right)\right\|_{\cJ_1}\qquad\qquad\qquad\qquad\qquad\qquad\qquad\qquad\qquad\nonumber\\
\leq
2^{2-\nu/4} (\pi
t)^{-\nu/2}\left\|e^{-t(H_0+2V_-)/2}\right\|_{\infty,\infty}^{1/2}\
\left\|e^{-t(H_0+4V_-)/4}\right\|_{\infty,\infty}^{1/2}\nonumber\\
\cdot\left(\|\chi_B\mathbf{P}_\bullet\{\tau_{B^c}\leq
t/2\}^{1/2}\|_{L^1}+\|\mathbf{E}_\bullet\{\chi_B(X_t);\tau_{B^c}\leq t/2
\}^{1/2}
\|_{L^1}\right),\label{St:gl:4}\\[5mm]
\left\|(1-\chi_B)\left( e^{-t(H_0+\chi_B V)}-e^{-t H_0}\right)\right\|_{\cJ_2}
\leq
2^{1-\nu/2}(\pi t)^{-\nu/4}\left\|e^{-t(H_0+2V_-)/2}
\right\|_{\infty,\infty}\nonumber\\ \cdot\left(3\|(1-\chi_B)\mathbf{P}_\bullet\{\tau_{B}\leq
t/2\}\|_{L^1}^{1/2}+ \|\mathbf{E}_\bullet\{1-\chi_B(X_t);\tau_{B}\leq t/2 \}
\|_{L^1}^{1/2}\right),\label{St:gl:5}\\[3mm]
\left\|(1-\chi_B)\left( e^{-t(H_0+\chi_B V)}-e^{-t H_0}\right)\right\|_{\cJ_1}\qquad\qquad\qquad\qquad\qquad\qquad\qquad\qquad\nonumber\\
\leq 2^{2-\nu/4} (\pi
t)^{-\nu/2}\left\|e^{-t(H_0+2V_-)/2}\right\|_{\infty,\infty}^{1/2}\
\left\|e^{-t(H_0+4V_-)/4}\right\|_{\infty,\infty}^{1/2}\nonumber\\
\cdot\left(\|(1-\chi_B)\mathbf{P}_\bullet\{\tau_{B}\leq
t/2\}^{1/2}\|_{L^1}+\|\mathbf{E}_\bullet\{1-\chi_B(X_t); \tau_B\leq t/2
\}^{1/2}\|_{L^1}\right).\label{St:gl:6}
\end{eqnarray}
\end{theorem}

This theorem can be also easily extended to the case where $H_0$ is replaced by
$H_0+W$ with $W$ being an arbitrary potential such that $W_+\in
K_\nu^{\mathrm{loc}}$ and $W_-\in K_\nu$.

\begin{lemma}\label{thm:Stollmann:3.1}
Let $B$ be an arbitrary domain in $\R^\nu$. Then for any $\epsilon>0$ there is
$C_\epsilon$ depending on $\epsilon$ only such that
\begin{eqnarray*}
\mathbf{E}_x\left\{\chi_B(X_t); \tau_{B^c}\leq t \right\}\leq
\left(\mathbf{P}_x\{\tau_{B^c}\leq t \} \right)^{1/2}\cdot\left\{
\begin{array}{ll}
1, & x\in B,\\ C_\epsilon \exp\left\{-\frac{\dist(x,B)^2}{2(4+\epsilon)t}
\right\}, & x\notin B
\end{array} \right.
\end{eqnarray*}
for all $t>0$.
\end{lemma}

\begin{proof}
By the Schwarz inequality with respect to the Wiener measure
\begin{eqnarray*}
\mathbf{E}_x\left\{\chi_B(X_t); \tau_{B^c}\leq t \right\}\leq
\left(\mathbf{E}_x\left\{\chi_B(X_t)\right\} \right)^{1/2}
\left(\mathbf{P}_x\{\tau_{B^c}\leq t \} \right)^{1/2}\\
=\left[\left(e^{-t H_0}\chi_B \right)(x) \right]^{1/2}
\left(\mathbf{P}_x\{\tau_{B^c}\leq t \} \right)^{1/2}.
\end{eqnarray*}
For $x\in B$ one obviously has $(e^{-tH_0}\chi_B)(x)\leq 1$. Now suppose that
$x\notin B$. Then
\begin{eqnarray*}
(e^{-tH_0}\chi_B)(x) &=& (4\pi t)^{-\nu/2}\int_B \exp\left\{-\frac{(x-y)^2}{4t}
\right\} dy\\ &=& (4\pi t)^{-\nu/2} \int_B
\exp\left\{-\frac{(x-y)^2}{(4+\epsilon)t}\right\}
\exp\left\{-\frac{\epsilon}{t}
\frac{(x-y)^2}{4\epsilon+16}\right\} dy\\
&\leq& (4\pi t)^{-\nu/2}
\sup_{y\in B}\left\{\exp\left\{-\frac{(x-y)^2}{(4+\epsilon)t}\right\}\right\}
\int_B \exp\left\{-\frac{\epsilon}{t}
\frac{(x-y)^2}{4\epsilon+16}\right\} dy\\
&\leq& (4\pi t)^{-\nu/2} \exp\left\{-\frac{\dist(x,B)^2}{(4+\epsilon)t}\right\}
\int_{\R^\nu} \exp\left\{-\frac{\epsilon}{t}
\frac{(x-y)^2}{4\epsilon+16}\right\} dy\\
&=&(4\pi)^{-\nu/2} \exp\left\{-\frac{\dist(x,B)^2}{(4+\epsilon)t}\right\}
\int_{\R^\nu} \exp\left\{-\epsilon
\frac{y^2}{4\epsilon+16}\right\} dy,
\end{eqnarray*}
which completes the proof of the lemma.
\end{proof}

Let $\meas(\cdot)$ denote the $\nu$-dimensional Lebesgue measure. Sometimes we
will make the dimensionality explicit and write $\meas_n$ for $1\leq n\leq
\nu$.

By Lemmas \ref{thm:Stollmann:2} and \ref{thm:Stollmann:3.1} as a corollary of
Theorem \ref{thm:Stollmann:3} we obtain

\begin{corollary}\label{thm:Stollmann:3.2}
Let $B$ be a box in $\R^\nu$. For $\nu\geq 2$ and for every $t>0$ there is
$c>0$ independent of $B$ such that
\begin{eqnarray}
\left\|\chi_B\left(e^{-t(H_0+V)}-e^{-t(H_0+\chi_B V)}\right) \right\|_{\cJ_2}&\leq&
c(\meas_{\nu-1}(\partial B))^{1/2},\label{St:gl:7}\\
\left\|\chi_B\left(e^{-t(H_0+V)}-e^{-t(H_0+\chi_B V)}\right)\right\|_{\cJ_1}&\leq&
c\ \meas_{\nu-1}(\partial B),\label{St:gl:8}\\
\left\|(1-\chi_B)\left( e^{-t(H_0+\chi_B V)}-e^{-t H_0}\right)\right\|_{\cJ_2}&\leq&
c(\meas_{\nu-1}(\partial B))^{1/2},\label{St:gl:9}\\
\left\|(1-\chi_B)\left( e^{-t(H_0+\chi_B V)}-e^{-t H_0}\right)\right\|_{\cJ_1}&\leq&
c\ \meas_{\nu-1}(\partial B).\label{St:gl:10}
\end{eqnarray}
If $\nu=1$ the same inequalities hold if the r.h.s. of (\ref{St:gl:7}) --
(\ref{St:gl:10}) are replaced by some constants.
\end{corollary}

Indeed to prove the corollary it suffices to estimate the integral of a
positive function ``concentrated" near the boundary $\partial B$ and falling
off exponentially fast away from $\partial B$. Lemmas \ref{thm:Stollmann:2} and
\ref{thm:Stollmann:3.1} say that the rate of fall-off depends only on $t$ and
the dimension $\nu$. Thus such integrals can be bounded by
$\meas_{\nu-1}(\partial B)$ times a constant depending on $t$ and $\nu$ only.

Actually Corollary \ref{thm:Stollmann:3.2} can be easily extended to more
complicated domains $\Lambda$. For instance we may consider the case where
there are two boxes $B_1$ and $B_2$, $B_1\subsetneq B_2$ such that
$\partial\Lambda\subset B_2\setminus B_1$. In this case Corollary
\ref{thm:Stollmann:3.2} is valid with $\meas_{\nu-1}(\partial B)$ on the
r.h.s.\ of
\eqref{St:gl:7} -- \eqref{St:gl:10} replaced by $\meas_\nu(B_2\setminus B_1)$.

We turn to the proof of Theorem \ref{thm:Stollmann:3}. By $H_0+V+\infty_B$ and
$H_0+V+\infty_{B^c}$ we denote the operator $H_0+V$ on $L^2(B^c)$ and $L^2(B)$
respectively with Dirichlet boundary conditions on $\partial B$. These
notations are motivated by the fact that the operators $H_0+V+\infty_B$ and
$H_0+V+\infty_{B^c}$ can be understood as limits of $H_0+V+k\chi_B$ and
$H_0+V+k\chi_{B^c}$ respectively as $k\rightarrow\infty$ (see e.g.\
\cite{Demuth:91}). Using the decomposition $L^2(\R^\nu)=L^2(B)\oplus L^2(B^c)$
these operators can be identified with the operators $0\oplus (H_0+V+\infty_B)$
and $(H_0+V+\infty_{B^c})\oplus 0$ acting on the whole $L^2(\R^\nu)$, so we
will use the same notations for these operators.

First we prove the following auxiliary inequalities
\begin{lemma}\label{thm:Stollmann:4}
Let $B\subset\R^\nu$ be a compact set. Let $V$ be a measurable function such
that $V_+\in K_\nu^{\mathrm{loc}}$ and $V_-\in K_\nu$. Then for any $t>0$
\begin{eqnarray}
&&\left\|\chi_B e^{-t(H_0+V)}-e^{-t(H_0+V+\infty_{B^c})}\right\|_{\cJ_2} \leq
(2\pi t)^{-\nu/4}\
\left\|e^{-t(H_0+2V_-)/2}\right\|_{\infty,\infty}\cdot \nonumber\\ &&\quad\cdot\left(
3\|\chi_B \mathbf{P}_\bullet\{\tau_{B^c}\leq t/2
\}\|^{1/2}_{L^1}+\|\mathbf{E}_{\bullet}\{\chi_B(X_t); \tau_{B^c}\leq t/2 \}
\|_{L^1}^{1/2} \right),\label{St:gl:11}\\
&&\left\|(1-\chi_B) e^{-t(H_0+\chi_B
V)}-e^{-t(H_0+\infty_{B})}\right\|_{\cJ_2}\leq (2\pi t)^{-\nu/4}\
\left\|e^{-t(H_0+2V_-)/2}\right\|_{\infty,\infty}\cdot
\nonumber\\&&
\quad\cdot\left( 3\|(1-\chi_B)
\mathbf{P}_\bullet\{\tau_B\leq t/2
\}\|^{1/2}_{L^1}+\|\mathbf{E}_{\bullet}\{1-\chi_B(X_t); \tau_{B}\leq t/2\}
\|^{1/2}_{L^1}\right).\label{St:gl:12}
\end{eqnarray}
\end{lemma}

\begin{proof}
First let us prove (\ref{St:gl:11}). We write the operator under the norm in
the form $\chi_B D(t)$ with
\begin{equation}\label{Dt.def}
D(t)= e^{-t(H_0+V)}-e^{-t(H_0+V+\infty_{B^c})}.
\end{equation}
By the semigroup property
\begin{eqnarray}\label{Dt.ref}
D(t)&=&e^{-t(H_0+V)/2}D(t/2)+D(t/2)e^{-t(H_0+V+\infty_{B^c})/2}\nonumber\\ &=&
D(t/2)^2 + D(t/2) e^{-t(H_0+V+\infty_{B^c})/2}+ e^{-t(H_0+V+\infty_{B^c})/2}
D(t/2),
\end{eqnarray}
and therefore
\begin{eqnarray}\label{St:gl:13}
\lefteqn{\|\chi_B D(t)\|_{\cJ_2}\leq \|\chi_B D(t/2)^2\|_{\cJ_2}\nonumber}\\
&+&\|\chi_B D(t/2) e^{-t(H_0+V+\infty_{B^c})/2}\|_{\cJ_2}\nonumber\\ &+&
\|D(t/2)\chi_B e^{-t(H_0+V+\infty_{B^c})/2}\|_{\cJ_2},
\end{eqnarray}
where we have used
$e^{-t(H_0+V+\infty_{B^c})}=e^{-t(H_0+V+\infty_{B^c})}\chi_B$ and the fact that
$\|A^\ast\|_{\cJ_p}=\|A\|_{\cJ_p}$. By the Feynman-Kac formula
\begin{eqnarray}\label{St:gl:14}
(D(t)f)(x) &=& \mathbf{E}_x\left\{\exp\left\{-\int_0^t V(X_s)ds
\right\}f(X_t)\right\}\nonumber\\
&& -\mathbf{E}_x\left\{\exp\left\{-\int_0^t V(X_s)ds
\right\}f(X_t);\ \tau_{B^c}\geq t\right\}\nonumber\\
&=&\mathbf{E}_x\left\{\exp\left\{-\int_0^t V(X_s)ds
\right\}f(X_t);\ \tau_{B^c}\leq t\right\}\geq 0
\end{eqnarray}
if $f\geq 0$. Thus $D(t)$ preserves positivity. The same is obviously valid for
the operator $e^{-t(H_0+V+\infty_{B^c})}$. Also $e^{-t(H_0+V)}$ and
$e^{-t(H_0+V+\infty_{B^c})}$ are bounded operators from $L^2$ to $L^\infty$
\cite{Simon:82}. Therefore we can apply Lemma \ref{Stollmann:HS} (see Appendix)
to estimate (\ref{St:gl:13}) thus obtaining
\begin{eqnarray*}
\|\chi_B D(t/2)^2\|_{\cJ_2} &\leq& \|\chi_B D(t/2)\|_{\infty,2}\ \|D(t/2)\|_{2,\infty},\\
\left\|\chi_B D(t/2) e^{-t(H_0+V+\infty_{B^c})/2}\right\|_{\cJ_2} &\leq&
\|\chi_B D(t/2)\|_{\infty,2}\ \left\|e^{-t(H_0+V+\infty_{B^c})/2}\right\|_{2,\infty},\\
\left\|D(t/2) \chi_B  e^{-t(H_0+V+\infty_{B^c})/2}\right\|_{\cJ_2} &\leq&
\|D(t/2)\chi_B \|_{\infty,2}\ \left\|e^{-t(H_0+V+\infty_{B^c})/2}\right\|_{2,\infty}.
\end{eqnarray*}
By Lemma \ref{A.3:neu}
\begin{displaymath}
\left\|e^{-t(H_0+V)/2} \right\|_{2,\infty}\leq (2\pi t)^{-\nu/4}
\left\|e^{-t(H_0+2V)/2} \right\|_{\infty,\infty}^{1/2}.
\end{displaymath}
By the monotonicity property (\ref{SSG2})
\begin{displaymath}
\left\|e^{-t(H_0+2V)/2} \right\|_{\infty,\infty}\leq
\left\|e^{-t(H_0+2V_-)/2} \right\|_{\infty,\infty}.
\end{displaymath}
Applying the Schwarz inequality with respect to the Wiener measure to the
Feynman-Kac formula we obtain
\begin{eqnarray*}
\lefteqn{\left|\left(e^{-t(H_0+V+\infty_{B^c})/2}f \right)(x) \right| =
\mathbf{E}_x\left\{\exp\left\{-2\int_0^{t/2}V(X_s)ds \right\}f(X_{t/2});
\tau_{B^c}> t/2 \right\}}\\
&\leq & \left(\mathbf{E}_x\left\{\exp\left\{-2\int_0^{t/2}V(X_s)ds
\right\}; \tau_{B^c}> t/2 \right\} \right)^{1/2}\
\left(\mathbf{E}_x\left\{ |f(X_{t/2})|^2\right\}
\right)^{1/2}\\
&\leq & \left(\mathbf{E}_x\left\{\exp\left\{-2\int_0^{t/2}V(X_s)ds
\right\} \right\} \right)^{1/2}\ \left(\mathbf{E}_x\left\{ |f(X_{t/2})|^2\right\}
\right)^{1/2}\\
&=& \left[\left(e^{-t(H_0+2V)/2}1 \right)(x) \right]^{1/2}\
\left[\left(e^{-tH_0/2}|f|^2
\right)(x)
\right]^{1/2}
\end{eqnarray*}
for any $f\in L^2$. This leads (see the proof of Lemma \ref{A.3:neu} in the
Appendix) to the inequality
\begin{equation}\label{nummer}
\left\|e^{-t(H_0+V+\infty_{B^c})/2} \right\|_{2,\infty}\leq (2\pi t)^{-\nu/4}
\left\|e^{-t(H_0+2V)/2} \right\|_{\infty,\infty}^{1/2},
\end{equation}
and thus
\begin{displaymath}
\|D(t/2)\|_{2,\infty}\leq
2(2\pi t)^{-\nu/4}\left\|e^{-t(H_0+2V)/2}\right\|_{\infty,\infty}^{1/2}.
\end{displaymath}
Now we estimate $\|\chi_B D(t/2)\|_{\infty,2}$. From the Feynman-Kac formula
(\ref{St:gl:14}) with $f\equiv 1$ by means of the Schwarz inequality with
respect to the Wiener measure we obtain
\begin{displaymath}
\left(D(t)1\right)(x)\leq \left(\mathbf{E}_x\left\{\exp\left\{-2\int_0^t V_-(X_s)ds
\right\} \right\} \right)^{1/2} \left(\mathbf{P}_x\{\tau_{B^c}\leq t \} \right)^{1/2},
\end{displaymath}
and hence
\begin{displaymath}
\|\chi_B D(t)\|_{\infty,2}\leq
\sup_x\left(\mathbf{E}_x\left\{\exp\left\{-2\int_0^t V_-(X_s)ds \right\}\right\}\right)^{1/2}
\|\chi_B \mathbf{P}_\bullet\{\tau_{B^c}\leq t \}\|_{L^1}^{1/2}.
\end{displaymath}
Now we note that
\begin{displaymath}
\sup_x\left(\mathbf{E}_x\left\{\exp\left\{-2\int_0^t V_-(X_s)ds \right\}\right\}\right)^{1/2}
= \left\|e^{-t(H_0+2 V_-)}1 \right\|_{L^\infty}^{1/2}=
\left\|e^{-t(H_0+2 V_-)} \right\|_{\infty,\infty}^{1/2}.
\end{displaymath}

We turn to the estimate of $\|D(t/2)\chi_B\|_{\infty,2}$. To this end we write
\begin{eqnarray*}
(D(t)\chi_B)(x) &=& \mathbf{E}_x\left\{\exp\left\{-\int_0^t V(X_s)ds
\right\}\chi_B(X_t);\ \tau_{B^c}\leq t\right\}\\
&\leq & \left(\mathbf{E}_x\left\{\exp\left\{-2\int_0^t V_-(X_s)ds\right\}
\right\} \right)^{1/2}\
\left(\mathbf{E}_x\left\{\chi_B(X_t);\ \tau_{B^c}\leq t) \right\} \right)^{1/2}\\
&\leq& \left\|e^{-t(H_0+2V_-)}\right\|_{\infty,\infty}^{1/2}\
\left(\mathbf{E}_x\left\{\chi_B(X_t);\ \tau_{B^c}\leq t) \right\} \right)^{1/2}.
\end{eqnarray*}
This completes the proof of the inequality (\ref{St:gl:11}).

The proof of (\ref{St:gl:12}) follows along the same lines. Denoting
\begin{displaymath}
D(t)= e^{-t(H_0+\chi_B V)} - e^{-t(H_0+\infty_B)}
\end{displaymath}
we obtain
\begin{eqnarray*}
D(t) &=& e^{-t(H_0+\chi_B V)/2} D(t/2) + D(t/2) e^{-t(H_0+\infty_B)/2}\\ &=&
D(t/2)^2 + D(t/2) e^{-t(H_0+\infty_B)/2}  + e^{-t(H_0+\infty_B)/2}  D(t/2),
\end{eqnarray*}
and therefore
\begin{eqnarray*}
\lefteqn{\|(1-\chi_B)D(t)\|_{\cJ_2}\leq \|(1-\chi_B)D(t/2)^2\|_{\cJ_2}}\\
&+&\left\|(1-\chi_B)D(t/2) e^{-t(H_0+\infty_B)/2} \right\|_{\cJ_2}\\ &+&\left\|
D(t/2)(1-\chi_B) e^{-t(H_0+\infty_B)/2}
\right\|_{\cJ_2}.
\end{eqnarray*}
Again by Lemma \ref{Stollmann:HS}
\begin{eqnarray*}
\|(1-\chi_B) D(t/2)^2\|_{\cJ_2} &\leq & \|(1-\chi_B)D(t/2)\|_{\infty,2} \\
&& \left(\left\|e^{-t(H_0+\chi_B V)/2}\right\|_{2,\infty}+
\left\|e^{-t(H_0+\infty_B )/2}\right\|_{2,\infty}\right),\\
\|(1-\chi_B)D(t/2)e^{-t(H_0+\infty_B)/2}\|_{\cJ_2} &\leq & \|(1-\chi_B)D(t/2)\|_{\infty,2}\
\left\| e^{-t(H_0+\infty_B)/2}\right\|_{2,\infty},\\
\|D(t/2)(1-\chi_B)e^{-t(H_0+\infty_B)/2}\|_{\cJ_2} &\leq& \|D(t/2)(1-\chi_B)\|_{\infty,2}\
\left\| e^{-t(H_0+\infty_B)/2}\right\|_{2,\infty}.
\end{eqnarray*}
By \eqref{nummer} and by the monotonicity property (\ref{SSG2})
\begin{displaymath}
\left\|e^{-t(H_0+\infty_B)/2}\right\|_{2,\infty}\leq (2\pi t)^{-\nu/4}
\left\|e^{-tH_0/2}\right\|^{1/2}_{\infty,\infty}\leq (2\pi t)^{-\nu/4}
\left\|e^{-t(H_0+2V_-)/2}\right\|^{1/2}_{\infty,\infty}.
\end{displaymath}
By Lemma \ref{A.3:neu} and again by the monotonicity property (\ref{SSG2})
\begin{displaymath}
\left\|e^{-t(H_0+\chi_B V)/2}\right\|_{2,\infty}\leq (2\pi t)^{-\nu/4}
\left\|e^{-t(H_0+2\chi_B V)/2}\right\|_{\infty,\infty}^{1/2}
\leq (2\pi t)^{-\nu/4}
\left\|e^{-t(H_0+2V_-)/2} \right\|_{\infty,\infty}^{1/2}.
\end{displaymath}
By the Feynman-Kac formula we obtain
\begin{displaymath}
\left(D(t)1\right)(x)\leq \left(\mathbf{E}_x\left\{\exp\left\{-2\int_0^t V_-(X_s)\chi_B(X_s)ds
\right\} \right\} \right)^{1/2} \left(\mathbf{P}_x\{\tau_{B}\leq t \} \right)^{1/2},
\end{displaymath}
which immediately gives
\begin{displaymath}
\|(1-\chi_B)D(t/2)\|_{\infty,2} \leq \left\|e^{-t(H_0+2V_-)/2}\right\|^{1/2}_{\infty,\infty}\
\|(1-\chi_B)\mathbf{P}_{\bullet}\{\tau_B\leq t/2\}\|_{L^1}^{1/2}.
\end{displaymath}
Further we consider
\begin{eqnarray*}
(D(t/2)(1-\chi_B))(x) &=& \mathbf{E}_x\left\{\exp\left\{-\int_0^{t/2}
V(X_s)ds\right\} (1-\chi_B(X_{t/2})); \tau_B\leq t/2
\right\}\\
&\leq& \left\|e^{-t(H_0+2V_-)/2}\right\|_{\infty,\infty}^{1/2}\
\left(\mathbf{E}_x\{1-\chi_B(X_{t/2}); \tau_B\leq t/2 \} \right)^{1/2},
\end{eqnarray*}
which completes the proof of (\ref{St:gl:12}).
\end{proof}

We are now in the position to prove the estimates (\ref{St:gl:3}) and
(\ref{St:gl:5}). We write
\begin{eqnarray*}
\chi_B e^{-t(H_0+V)}-\chi_B e^{-t(H_0+\chi_B V)} &=&
\chi_B e^{-t(H_0+V)} - e^{-t(H_0+V+\infty_{B^c})}\\
&& -\left(\chi_B e^{-t(H_0+\chi_B V)} - e^{-t(H_0+\chi_B
V+\infty_{B^c})}\right)
\end{eqnarray*}
and apply Lemma \ref{thm:Stollmann:4}. This gives (\ref{St:gl:3}). Similarly we
obtain (\ref{St:gl:5}).

We turn now to the trace class estimates (\ref{St:gl:4}) and (\ref{St:gl:6}).
As in the Hilbert-Schmidt case we start with an auxiliary lemma:

\begin{lemma}\label{thm:Stollmann:5}
Let $B\subset\R^\nu$ be a compact set. Let $V$ be a measurable function such
that $V_+\in K_\nu^{\mathrm{loc}}$ and $V_-\in K_\nu$. Then for any $t>0$
\begin{eqnarray}
\left\|\chi_B e^{-t(H_0+V)}-e^{-t(H_0+V+\infty_{B^c})}\right\|_{\cJ_1}\qquad\qquad\qquad\qquad\qquad\qquad\qquad\qquad\nonumber\\ \leq
2^{1-\nu/4}(\pi
t)^{-\nu/2}\left\|e^{-t(H_0+2V_-)/2}\right\|_{\infty,\infty}^{1/2}\
\left\|e^{-t(H_0+4V_-)/4}\right\|_{\infty,\infty}^{1/2}\nonumber\\
\cdot\left(\|\chi_B \mathbf{P}_\bullet\{\tau_{B^c}\leq t/2
\}^{1/2}\|_{L^1}+\|\mathbf{E}_{\bullet}\{\chi_B (X_t); \tau_{B^c}\leq t/2\}^{1/2} \|_{L^1}\right),\label{St:gl:15}\\
\qquad\left\|(1-\chi_B) e^{-t(H_0+\chi_B V)}-e^{-t(H_0+V+\infty_{B})}\right\|_{\cJ_1}\qquad\qquad\qquad\qquad\qquad\qquad\nonumber\\
\leq 2^{1-\nu/4}(\pi
t)^{-\nu/2}\left\|e^{-t(H_0+2V_-)/2}\right\|_{\infty,\infty}^{1/2}\
\left\|e^{-t(H_0+4V_-)/4}\right\|_{\infty,\infty}^{1/2}\nonumber\\
\cdot\left(\|(1-\chi_B) \mathbf{P}_\bullet\{\tau_B\leq t/2
\}^{1/2}\|_{L^1}+\|\mathbf{E}_{\bullet}\{1-\chi_B (X_t); \tau_{B}\leq t/2\}^{1/2} \|_{L^1}\right).\label{St:gl:16}
\end{eqnarray}
\end{lemma}

\begin{proof}
We prove (\ref{St:gl:15}) only since the proof of (\ref{St:gl:16}) follows
along the same lines. Again we use the representation of the operator under the
norm in the form $\chi_B D(t)$ with $D(t)$ being defined by (\ref{Dt.def}). By
means of the identity (\ref{Dt.ref}) we estimate
\begin{eqnarray}\label{Dt.next}
\|\chi_B D(t)\|_{\cJ_1}&\leq& \|D(t/2)^2\chi_B\|_{\cJ_1}\nonumber\\
&&+\|e^{-t(H_0+V+\infty_{B^c})/2}
\chi_B D(t/2)\chi_B\|_{\cJ_1}\nonumber\\
&& +\|e^{-t(H_0+V+\infty_{B^c})/2} \chi_B D(t/2)\|_{\cJ_1}.
\end{eqnarray}
Choose  an arbitrary $f\in L^2(\R^\nu)$ with $\|f\|_{L^2}\leq 1$ and consider
\begin{eqnarray*}
\lefteqn{|(D(t)f)(x)|=\left|\mathbf{E}_x\left\{\exp\left\{-\int_0^t
V(X_s)ds\right\}f(X_t); \tau_{B^c}\leq t
\right\}\right|}\\
&\leq& \left(\mathbf{E}_x\left\{\exp\left\{-2\int_0^t
V_-(X_s)ds\right\}|f(X_t)|^2 \right\}\right)^{1/2}\
\left(\mathbf{P}_x\left\{\tau_{B^c}\leq t\right\}
\right)^{1/2}\\
&\leq&\sup_f \sup_x \left(\mathbf{E}_x\left\{\exp\left\{-2\int_0^t
V_-(X_s)ds\right\}|f(X_t)|^2 \right\}\right)^{1/2}\
\left(\mathbf{P}_x\left\{\tau_{B^c}\leq t\right\}
\right)^{1/2}\\
&=&\left\|e^{-t(H_0+2V_-)}\right\|_{1,\infty}^{1/2}\
\left(\mathbf{P}_x\left\{\tau_{B^c}\leq t\right\}
\right)^{1/2}.
\end{eqnarray*}
Similarly we have
\begin{eqnarray*}
\lefteqn{|(D(t)\chi_B f)(x)| = \left|\mathbf{E}_x\left\{\exp\left\{-\int_0^t
V(X_s)ds\right\}\chi_B(X_t)f(X_t); \tau_{B^c}\leq t
\right\} \right|}\\
&\leq & \left(\mathbf{E}_x\left\{\exp\left\{-2\int_0^t
V_-(X_s)ds\right\}|f(X_t)|^2 \right\}\right)^{1/2}\
\left(\mathbf{E}_x\left\{\chi_B(X_t); \tau_{B^c}\leq t\right\}
\right)^{1/2}\\
&\leq& \left\|e^{-t(H_0+2V_-)}\right\|_{1,\infty}^{1/2}\
\left(\mathbf{E}_x\left\{\chi_B(X_t); \tau_{B^c}\leq t\right\}
\right)^{1/2}.
\end{eqnarray*}
Since $D(t)$ preserves positivity and $e^{-t(H_0+V)}$,
$e^{-t(H_0+V+\infty_{B^c})}$ are bounded as maps from $L^1$ to $L^2$
\cite{Simon:82} we can use Lemma \ref{Stollmann:Tr} to estimate
(\ref{Dt.next}), which immediately leads to
\begin{eqnarray*}
\|\chi_B D(t)\|_{\cJ_1} \leq \|D(t/2)\|_{1,2}\ \left\|e^{-t(H_0+2V_-)/2}\right\|^{1/2}_{1,\infty}\
\|\mathbf{E}_\bullet \{\chi_B(X_t): \tau_{B^c}\leq t/2 \}^{1/2}\|_{L^1}\\
+2\left\|e^{-t(H_0+V+\infty_{B^c})/2} \right\|_{1,2}\
\left\|e^{-t(H_0+2V_-)/2}\right\|^{1/2}_{1,\infty}\
\|\chi_B \mathbf{P}_\bullet\{\tau_{B^c}\leq t/2\}^{1/2}\|_{L^1}.
\end{eqnarray*}
Since $e^{-t(H_0+V)}$ is self-adjoint, by duality (see e.g. \cite{Simon:82}) we
have
\begin{equation}\label{duality}
\left\|e^{-t(H_0+V)/2} \right\|_{1,2} = \left\|e^{-t(H_0+V)/2} \right\|_{2,\infty}.
\end{equation}
Applying Lemma \ref{A.3:neu} we obtain
\begin{displaymath}
\left\|e^{-t(H_0+V)/2} \right\|_{1,2}\leq (2\pi t)^{-\nu/4}
\left\|e^{-t(H_0+2V)/2} \right\|^{1/2}_{\infty,\infty}.
\end{displaymath}
From (\ref{duality}), (\ref{nummer}) and the monotonicity of the norm
(\ref{SSG2}) it follows that
\begin{displaymath}
\left\|e^{-t(H_0+V+\infty_{B^c})/2} \right\|_{1,2}\leq (2\pi t)^{-\nu/4}
\left\|e^{-t(H_0+2V)/2} \right\|_{\infty,\infty}^{1/2}\leq (2\pi t)^{-\nu/4}
\left\|e^{-t(H_0+2V_-)/2} \right\|_{\infty,\infty}^{1/2}.
\end{displaymath}
By the semigroup property and by \eqref{duality}
\begin{displaymath}
\left\|e^{-t(H_0+2V_-)/2} \right\|_{1,\infty}\leq \left\|e^{-t(H_0+2V_-)/4} \right\|_{1,2}\
\left\|e^{-t(H_0+2V_-)/4} \right\|_{2,\infty} = \left\|e^{-t(H_0+2V_-)/4} \right\|_{2,\infty}^2.
\end{displaymath}
Applying now Lemma \ref{A.3:neu} to the r.h.s. of this inequality we obtain
\begin{displaymath}
\left\|e^{-t(H_0+2V_-)/2} \right\|_{1,\infty}\leq (\pi t)^{-\nu/2}
\left\|e^{-t(H_0+4V_-)/4} \right\|_{\infty,\infty},
\end{displaymath}
thus completing the proof of (\ref{St:gl:15}).
\end{proof}

Similar to the case of the Hilbert-Schmidt norm this lemma immediately yields
(\ref{St:gl:4}) and (\ref{St:gl:6}).

Now we can prove the statements formulated in the Introduction (equations
(\ref{DOS.4}) and (\ref{DOS.6})):

\begin{theorem}\label{th.SSD.1}
Let $V$ be such that $V_+\in K_\nu^{\mathrm{loc}}$ and $V_-\in K_\nu$. Then for
any $g\in C_0^2$ and any sequence of boxes $\Lambda$ tending to infinity
\begin{displaymath}
\lim_{\Lambda\rightarrow\infty}(\meas(\Lambda))^{-1}
\tr\left[\chi_\Lambda\left(g(H_0+V)-g(H_0+\chi_\Lambda V)\right)
\right]=0,
\end{displaymath}
and
\begin{displaymath}
\lim_{\Lambda\rightarrow\infty}(\meas(\Lambda))^{-1}
\tr\left[(1-\chi_\Lambda)(g(H_0+\chi_\Lambda V)-g(H_0))\right]=0.
\end{displaymath}
\end{theorem}

\begin{proof}
Given $g\in C_0^2$ by the Stone--Weierstrass theorem we can find polynomials
$P_k(\lambda)$ in $e^{-\lambda}$ such that
\begin{displaymath}
\sup_{\lambda\in A}e^\lambda\left|g(\lambda)-P_k(\lambda)\right|\rightarrow 0,\quad
\sup_{\lambda\in A}e^\lambda\left|g'(\lambda)-P'_k(\lambda)\right|\rightarrow 0,
\quad A=\bigcup_\Lambda\spec(H_0+V_\Lambda),
\end{displaymath}
as $k\rightarrow\infty$ (see \cite{Simon:82}). Indeed, denoting
$x=e^{-\lambda}\in (0,\exp(-\inf A)]$ and $\widetilde{g}(x)=g(-\log x)$ we can
find polynomials $P_k(x)$ such that
\begin{equation}\label{Stone:Weierstrass}
\sup_x|\widetilde{g}(x)-P_k(x)|\rightarrow 0,\quad
\sup_x|\widetilde{g}'(x)-P'_k(x)|\rightarrow 0,\quad
\sup_x|\widetilde{g}''(x)-P''_k(x)|\rightarrow 0
\end{equation}
and $P_k(0)=P'_k(0)=0$. Since $\inf\spec(H_0+V_\Lambda)$ depends on the Kato
norm of $V_\Lambda$ only, the set $A$ is bounded below. Let $x_0$ be such that
$0<x_0<\inf\supp\ \widetilde{g}$. By \eqref{Stone:Weierstrass}
\begin{displaymath}
\sup_{x\geq x_0}\frac{|\widetilde{g}(x)-P_k(x)|}{x}\rightarrow 0, \quad
\sup_{x\geq x_0}\frac{|\widetilde{g}'(x)-P'_k(x)|}{x}\rightarrow 0
\end{displaymath}
as $k\rightarrow\infty$. For $x\in[0,x_0]$ by the mean value theorem we have
\begin{eqnarray*}
\sup_{x\in[0,x_0]}\frac{|\widetilde{g}(x)-P_k(x)|}{x}=\sup_{x\in[0,x_0]}\frac{P_k(x)}{x}
\leq \sup_{x\in[0,x_0]} |P'_k(x)|\rightarrow 0,\\
\sup_{x\in[0,x_0]}\frac{|\widetilde{g}'(x)-P'_k(x)|}{x}=\sup_{x\in[0,x_0]}\frac{P'_k(x)}{x}
\leq \sup_{x\in[0,x_0]} |P''_k(x)|\rightarrow 0
\end{eqnarray*}
as $k\rightarrow\infty$.

Let $F_k(\lambda)=e^\lambda[g(\lambda)-P_k(\lambda)]$. Obviously
\begin{eqnarray}\label{add.neu.1}
\tr\left[\chi_\Lambda(g(H_0+V)-g(H_0+\chi_\Lambda V))\right] &=&
\tr\left[\chi_\Lambda(g(H_0+V)-P_k(H_0+V)) \right]\nonumber\\
&&-\tr\left[\chi_\Lambda(g(H_0+\chi_\Lambda V)
-P_k(H_0+\chi_\Lambda V))\right]\nonumber\\
&&+\tr\left[\chi_\Lambda(P_k(H_0+V)-P_k(H_0+\chi_\Lambda V)) \right].
\end{eqnarray}
Then
\begin{eqnarray*}
\lefteqn{\left|\tr\left(\chi_\Lambda g(H_0+V)\chi_\Lambda-
\chi_\Lambda P_k(H_0+V)\chi_\Lambda \right)\right|}\\&=&
\left|\tr\left(\chi_\Lambda e^{-(H_0+V)/2}F_k(H_0+V)e^{-(H_0+V)/2}\chi_\Lambda\right)\right|\\
&\leq&\|F_k\|_{L^\infty}\tr(\chi_\Lambda e^{-(H_0+V)}\chi_\Lambda),
\end{eqnarray*}
\begin{eqnarray*}
\lefteqn{\left|\tr\left(\chi_\Lambda g(H_0+\chi_\Lambda V)\chi_\Lambda-
\chi_\Lambda P_k(H_0+\chi_\Lambda V)\chi_\Lambda)\right)\right|}\\&=&
\left|\tr\left(\chi_\Lambda e^{-(H_0+\chi_\Lambda V)/2}F_k(H_0+\chi_\Lambda V)
e^{-(H_0+\chi_\Lambda V)/2} \right)\right|\\ &\leq&
\|F_k\|_{L^\infty}\tr\left(\chi_\Lambda e^{-(H_0+\chi_\Lambda
V)}\chi_\Lambda\right).
\end{eqnarray*}
Dividing these inequalities by $\meas(\Lambda)$ and taking the limit
$\Lambda\rightarrow\infty$ gives
\begin{eqnarray*}
\ulim_{\Lambda\rightarrow\infty}(\meas(\Lambda))^{-1}
|\tr\left[\chi_\Lambda(g(H_0+V)-P_k(H_0+V)) \right]| &\leq& C
\|F_k\|_{L^\infty},\\
\ulim_{\Lambda\rightarrow\infty}(\meas(\Lambda))^{-1}
|\tr\left[\chi_\Lambda(g(H_0+\chi_\Lambda V)-P_k(H_0+\chi_\Lambda V))
\right]| &\leq& C
\|F_k\|_{L^\infty}
\end{eqnarray*}
with an appropriate constant $C>0$ independent of $k$. The third term on the
r.h.s.\ of
\eqref{add.neu.1} can be written in the form
\begin{displaymath}
\sum_{j=1}^k a_j \tr\left[\chi_\Lambda\left( e^{-j(H_0+V)}-e^{-j(H_0+\chi_\Lambda V)}
\right)\right]
\end{displaymath}
with $a_j$ being the coefficients of $P_k(\lambda)$, and thus by Corollary
\ref{thm:Stollmann:3.2}
\begin{displaymath}
\lim_{\Lambda\rightarrow\infty}(\meas(\Lambda))^{-1}
\tr[\chi_\Lambda(P_k(H_0+V)-P_k(H_0+\chi_\Lambda V))]=0
\end{displaymath}
for any $k$. We have proved that
\begin{displaymath}
\ulim_{\Lambda\rightarrow\infty}(\meas(\Lambda))^{-1}
|\tr[\chi_\Lambda(g(H_0+V)-g(H_0+\chi_\Lambda V))]|\leq 2C \|F_k\|_{L^\infty}
\end{displaymath}
for any $k\in\N$. Taking the limit $k\rightarrow\infty$ proves the first part
of the claim.

To prove the second part we write
\begin{eqnarray}\label{add.neu.2}
\lefteqn{\tr[(1-\chi_\Lambda)(g(H_0+\chi_\Lambda V)-g(H_0))]}\nonumber\\
&=& \tr[g(H_0+\chi_\Lambda V)-P_k(H_0+\chi_\Lambda
V)-g(H_0)+P_k(H_0)]\nonumber\\ &&
-\tr[\chi_\Lambda(g(H_0+\chi_\Lambda V)-P_k(H_0+\chi_\Lambda V))]\nonumber\\
&&+\tr[\chi_\Lambda(g(H_0)-P_k(H_0))]+\tr[(1-\chi_\Lambda)(P_k(H_0+\chi_\Lambda)-
P_k(H_0))].
\end{eqnarray}
Here the second and third terms can be considered as above thus giving
\begin{eqnarray*}
\ulim_{\Lambda\rightarrow\infty}(\meas(\Lambda))^{-1}
|\tr\left[\chi_\Lambda(g(H_0+\chi_\Lambda V)-P_k(H_0+\chi_\Lambda V)) \right]|
&\leq& C
\|F_k\|_{L^\infty},\\
\ulim_{\Lambda\rightarrow\infty}(\meas(\Lambda))^{-1}
|\tr\left[\chi_\Lambda(g(H_0)-P_k(H_0))
\right]| &\leq& C
\|F_k\|_{L^\infty}
\end{eqnarray*}
with an appropriate constant $C>0$. The fourth term divided by $\meas(\Lambda)$
by Corollary \ref{thm:Stollmann:3.2} tends to zero as
$\Lambda\rightarrow\infty$ for any $k\in\N$. Let
$\widetilde{F}_k(\lambda)=g(\lambda)-P_k(\lambda)$. By assumption
$\widetilde{F}\in C^2$. We write now the first term on the r.h.s.\ of
\eqref{add.neu.2} in the form
\begin{eqnarray*}
\lefteqn{\tr[g(H_0+\chi_\Lambda V)-P_k(H_0+\chi_\Lambda V)-g(H_0)+P_k(H_0)]}\\
&=&-\int_\R \widetilde{F}'_k(\lambda) \xi(\lambda; H_0+\chi_\Lambda V,
H_0)d\lambda,
\end{eqnarray*}
where $\xi(\lambda; H_0+\chi_\Lambda V, H_0)$ is the spectral shift function
for the pair of operators ($H_0+\chi_\Lambda V$, $H_0$). It can be constructed
from the spectral shift function for the pair ($e^{-t(H_0+\chi_\Lambda V)}$,
$e^{-tH_0}$) by means of the invariance principle. Thus the absolute value of
the first term on the r.h.s.\ of
\eqref{add.neu.2} can be bounded by
\begin{eqnarray*}
\int_\R |\widetilde{F}'_k(\lambda)| |\xi(\lambda; H_0+\chi_\Lambda V,
H_0)| d\lambda = \int_\R |e^\lambda \widetilde{F}'_k(\lambda)| e^{-\lambda}
|\xi(\lambda; H_0+\chi_\Lambda V, H_0)| d\lambda\\
\leq \sup_{\lambda\in A}|e^\lambda \widetilde{F}'_k(\lambda)|\int_\R
e^{-\lambda} |\xi(\lambda; H_0+\chi_\Lambda V, H_0)| d\lambda \leq
\sup_{\lambda\in A}|e^\lambda \widetilde{F}'_k(\lambda)|
\left\|e^{-(H_0+\chi_\Lambda V)}-e^{-H_0} \right\|_{\cJ_1}.
\end{eqnarray*}
By Theorem \ref{thm:Stollmann:1} and Lemma \ref{thm:Stollmann:2} it follows
that for any $k\in\N$
\begin{eqnarray*}
\ulim_{\Lambda\rightarrow\infty}(\meas(\Lambda))^{-1}
|\tr[g(H_0+\chi_\Lambda V)-P_k(H_0+\chi_\Lambda V)-g(H_0)+P_k(H_0)]|\\ \leq C
\sup_{\lambda\in A}|e^{\lambda}\widetilde{F}'_k(\lambda)|
\end{eqnarray*}
with some constant $C>0$ independent of $k$. Taking the limit
$k\rightarrow\infty$ completes the proof.
\end{proof}

\begin{corollary}\label{th.SSD.3}
If the density of states measure exists, then for any $g\in C_0^2$ and any
sequence of boxes $\Lambda$ tending to infinity
\begin{eqnarray}\label{St:gl:17}
\mu(g)-\mu_0(g) &=& \lim_{\Lambda\rightarrow\infty}
(\meas(\Lambda))^{-1}\tr\left[g(H_0+\chi_\Lambda V)-g(H_0)\right]\nonumber\\
&=& \lim_{\Lambda\rightarrow\infty} (\meas(\Lambda))^{-1}
\int_\R g'(\lambda) \xi(\lambda;H_0+\chi_\Lambda V,H_0) d\lambda.
\end{eqnarray}
Conversely, if the limit on the r.h.s.\ of (\ref{St:gl:17}) exists then also
the density of states measure exists and the equality (\ref{St:gl:17}) holds.
\end{corollary}

\begin{remark}
Actually in the formulation of Theorem \ref{th.SSD.1} and Corollary
\ref{th.SSD.3} instead of a sequence of boxes $\Lambda$ we can take a sequence
of arbitrary domains with piecewise smooth boundary tending to infinity in the
sense of Fisher.
\end{remark}

Before we complete this section we mention one more consequence of Lemma
\ref{thm:Stollmann:5}. Let $H=H_0+V$ with $V_+\in K_\nu^{\mathrm{loc}}$ and
$V_-\in K_\nu$. For an arbitrary bounded open set $B$ denote
$H_B^{(D)}=(H+\infty_B)\oplus(H+\infty_{B^c})$.

\begin{corollary}\label{th.SSD.3.1}
For any $t>0$
\begin{eqnarray}\label{St:gl:18}
\left\|e^{-tH}-e^{-tH_B^{(D)}}\right\|_{\cJ_1}\leq
2^{2-\nu/4}(\pi
t)^{-\nu/2}\left\|e^{-t(H_0+2V_-)/2}\right\|_{\infty,\infty}^{1/2}\
\left\|e^{-t(H_0+4V_-)/4}\right\|^{1/2}_{\infty,\infty}\nonumber\\
\cdot\Big(\|\chi_B \mathbf{P}_\bullet\{\tau_{B^c}\leq t/2\}^{1/2}\|_{L^1}+
\|(1-\chi_B)\mathbf{P}_\bullet \{\tau_B\leq t/2 \}^{1/2}\|_{L^1}\nonumber\\
+\|\mathbf{E}_\bullet\{\chi_B(X_t); \tau_{B^c}\leq t/2 \}^{1/2} \|_{L^1} +
\|\mathbf{E}_\bullet\{1-\chi_B(X_t); \tau_{B}\leq t/2 \}^{1/2} \|_{L^1}\Big).
\end{eqnarray}
\end{corollary}

\begin{proof}
We estimate
\begin{eqnarray*}
\left\|e^{-tH}-e^{-tH_B^{(D)}}\right\|_{\cJ_1}\leq
\left\|\chi_B e^{-tH}-e^{-t(H+\infty_{B^c})}\right\|_{\cJ_1}+
\left\|(1-\chi_B) e^{-tH}-e^{-t(H+\infty_{B})}\right\|_{\cJ_1}
\end{eqnarray*}
and apply Lemma \ref{thm:Stollmann:5}.
\end{proof}

If $B$ is a domain with convex boundary (e.g. a box or a ball) by means of
Lemmas \ref{thm:Stollmann:2} and \ref{thm:Stollmann:3.1} the expression in the
brackets in (\ref{St:gl:18}) can be bounded by $\meas_{\nu-1}(\partial B)$. Let
us fix some $E> -\inf\spec(H)\geq 0$. Due to the operator identity
\begin{displaymath}
(H+E)^{-m}=\frac{1}{\Gamma(m)}\int_0^\infty e^{-tH} e^{-tE} t^{m-1} dt
\end{displaymath}
for all $m>\nu/2$ one can easily obtain the estimate
\begin{displaymath}
\left\|(H+E)^{-m}-(H^{(D)}_B+E)^{-m}\right\|_{\cJ_1}\leq C \meas_{\nu-1}(\partial B).
\end{displaymath}
Inequalities of this type were studied earlier by Alama, Deift and Hempel
\cite{Alama:Deift:Hempel} and by Hempel \cite{Hempel:92}.

%%%%%%%%%%%%%%%%%%%%%%%%%%%%%%%%%%%%%%
\section{Lattices of Potentials}\label{sec:lattices}
%%%%%%%%%%%%%%%%%%%%%%%%%%%%%%%%%%%%%%
\setcounter{equation}{0}

Let $\L=\L^\nu=\{x_{\mathbf{j}}\}_{\mathbf{j}\in\Z^\nu}$ be a lattice in
$\R^\nu$ with basis $\{a_{k}\}_{k=1}^{\nu}$, i.e. every $x_{\mathbf{j}}$ can be
uniquely represented in the form $x_{\mathbf{j}}=a_1 j_1+\ldots+a_\nu j_\nu$
with some $\mathbf{j}=(j_1,\ldots,j_\nu)\in\Z^\nu$. With this lattice we
associate the
\textit{Birman-Solomyak class} $l^q(L^p;\L)$, which is the linear space of all
measurable functions for which the norm
\begin{displaymath}
\|f\|_{l^q(L^p;\L^\nu)} = \left(\sum_{j\in\Z\nu}\left[\int_{\Delta_{\mathbf{j}}^{\L^\nu}}
|f(x)|^p dx \right]^{q/p} \right)^{1/q},
\end{displaymath}
is finite. Here $\Delta_{\mathbf{j}}^{\L}$ is an elementary cell in $\R^\nu$
defined by $\L$ and centered at $x=x_{\mathbf{j}}$. In the case $\L=\Z^\nu$ we
have $l^q(L^p;\L)=l^q(L^p)$, the standard Birman-Solomyak class
\cite{Birman:Solomyak:69,Simon:82} associated with the integer lattice $\Z^\nu$
\begin{displaymath}
l^q(L^p)\equiv l^q(L^p;\Z^\nu)=\left\{f\Big|\
\|f\|_{l^q(L^p)}=\left(\sum_{\mathbf{j}\in{\Z}^\nu}\Big[\int_{\Delta_{\mathbf{j}}}
|f(x)|^pdx\Big]^{q/p}\right)^{1/q}<\infty\right\},
\end{displaymath}
where $\Delta_{\mathbf{j}}$ are unit cubes with centers at $x=\mathbf{j}$. In
particular, $l^1(L^2)\subset L^1(\R^\nu)\cap L^2(\R^\nu)$ for all $\nu$. It is
easy to see that the norms corresponding to different $\L$'s are equivalent,
i.e. for arbitrary lattices $\L_1$ and $\L_2$ of the above form there is
$0<c<1$ such that
\begin{displaymath}
c\|f\|_{l^q(L^p;\L_1)}\leq \|f\|_{l^q(L^p;\L_2)}\leq
c^{-1}\|f\|_{l^q(L^p;\L_1)}
\end{displaymath}
for all $f\in l^q(L^p)$.

Here we will consider potentials having the form
\begin{equation}\label{v.def}
V(x)=\sum_{\mathbf{j}\in \Z^\nu} f_{\mathbf{j}}(x-x_{\mathbf{j}}),
\end{equation}
where $x_{\mathbf{j}}\in \L^\nu$ and $f_{\mathbf{j}}$ is a family of
real-valued functions which are in the Birman-Solomyak class $l^1(L^2)$
uniformly, i.e.
\begin{equation}\label{lat.1}
\sup_{\mathbf{j}\in \Z^\nu} \|f_{\mathbf{j}}\|_{l^1(L^2)}<\infty,\quad
\sum_{\mathbf{j}\in\Z^\nu} \sup_{\mathbf{k}\in\Z^\nu}\|\chi_{\Delta_{\mathbf{j}}^{\L^\nu}}
f_{\mathbf{k}}\|_{L^2}<\infty,
\end{equation}
and if $\nu\geq 4$ in addition uniformly in $L^p$ for some $p>\nu/2$, i.e.
\begin{equation}\label{lat.2}
\sup_{\mathbf{j}\in \Z^\nu} \|f_{\mathbf{j}}\|_{L^p}<\infty.
\end{equation}

Under the conditions (\ref{lat.1}), (\ref{lat.2}) the potential $V$ is in
$L^1_{\mathrm{unif,loc}}(\R^\nu)\cap L^2_{\mathrm{unif,loc}}(\R^\nu)$ for
$\nu\leq 3$ and in $L^1_{\mathrm{unif,loc}}(\R^\nu)\cap
L^p_{\mathrm{unif,loc}}(\R^\nu)$ for some $p>\nu/2$ if $\nu\geq 4$. (Recall
that $V\in L^p_{\mathrm{unif,loc}}(\R^\nu)$ iff $\sup_y\int_{|x-y|\leq 1}
|V(x)|^p dx <\infty$). Thus $V\in K_\nu$ and therefore $H=H_0+V$ is defined in
the form sense with $\cQ(H)=\cQ(H_0)$ and is self-adjoint.

Denote
\begin{displaymath}
V_\Lambda=\sum_{\substack{\mathbf{j}\in\L \\
\mathbf{j}\in\Lambda}}f_{\mathbf{j}}(\cdot-x_{\mathbf{j}})
\end{displaymath}
such that $V_\Lambda\rightarrow V$ a.e.\ as $\Lambda\rightarrow\infty$. Now we
formulate the main result of the present section:

\begin{theorem}\label{th.SSD:main}
Let the potential $V$ will be given by (\ref{v.def}) such that (\ref{lat.1})
and (\ref{lat.2}) are fulfilled. Then for any $g\in C_0^2$ and any sequence of
boxes $\Lambda$ tending to infinity
\begin{displaymath}
\lim_{\Lambda\rightarrow\infty} (\meas(\Lambda))^{-1}
\tr[g(H_0+\chi_\Lambda V)-g(H_0+V_\Lambda)]=0.
\end{displaymath}
\end{theorem}

As above instead of boxes we can take a sequence of arbitrary domains with
piecewise smooth boundary tending to infinity in the sense of Fisher. We start
the proof with the following

\begin{lemma}\label{th.SSD.4}
Let $V_1,V_2$ be such that $(V_i)_+\in K_\nu^{\mathrm{loc}}$, $(V_i)_-\in
K_\nu$, $i=1,2$ and $V_1-V_2\in l^1(L^2)$. Then for all $t>0$ there is a
constant $C_t$ depending on $t$ only such that
\begin{eqnarray}\label{lat.3}
\lefteqn{\left\|e^{-t(H_0+V_1)}-e^{-t(H_0+V_2)}\right\|_{\cJ_1}}\nonumber\\ && \leq
C_t \sup_{\tau\in(0,t)}\left\|e^{-\tau(H_0+V_1)/2}\right\|_{2,2}\
\sup_{\tau\in(0,t)}\left\|e^{-\tau(H_0+V_2)/2}\right\|_{2,2}\cdot\nonumber\\
&&\cdot\left\|e^{-t(H_0+2V_1)/2}\right\|_{1,\infty}\
\left\|e^{-t(H_0+2V_2)/2}\right\|_{1,\infty}\
\|V_1-V_2\|_{l^1(L^2)}.
\end{eqnarray}
\end{lemma}

\begin{proof}
The proof of that $V_1-V_2\in l^1(L^2)$ implies
$\exp\{-t(H_0+V_1)\}-\exp\{-t(H_0+V_2)\}$ is trace class was given by Simon
\cite{Simon:79b,Simon:82}. To obtain the estimate (\ref{lat.3}) we simply
repeat the arguments of Simon explicitly controlling the constants in the
intermediate estimates.

We make use of the DuHamel formula and write
\begin{eqnarray*}
\lefteqn{e^{-t(H_0+V_1)}-e^{-t(H_0+V_2)}=\int_0^t ds\ e^{-s(H_0+V_1)}(V_1-V_2)
e^{-(t-s)(H_0+V_2)}}\\ &=& \int_0^{t/2} ds\
e^{-s(H_0+V_1)}(V_1-V_2)e^{-(t-s)(H_0+V_2)}\\&&+
\int_{t/2}^t ds\ e^{-s(H_0+V_1)}(V_1-V_2)
e^{-(t-s)(H_0+V_2)} \\ &=& \frac{t}{2}
\int_0^1 d\tau\ e^{-t\tau(H_0+V_1)/2}(V_1-V_2)e^{-t(H_0+V_2)/2}e^{-t(1-\tau)(H_0+V_2)/2}\\
&& + \frac{t}{2}\int_0^1 d\tau\
e^{-t\tau(H_0+V_1)/2}e^{-t(H_0+V_1)/2}(V_1-V_2)e^{-t(1-\tau)(H_0+V_2)/2},
\end{eqnarray*}
which holds initially weakly. However, by means of the estimate (\ref{SSG3})
with $p=q=2$ and the fact that $(V_1-V_2)e^{-t(H_0+V_2)}$ and
$e^{-t(H_0+V_1)}(V_1-V_2)$ are trace class \cite[Theorem B.9.2]{Simon:82} this
identity can be seen to hold in the trace norm sense. Therefore we obtain
\begin{eqnarray*}
\lefteqn{\left\|e^{-t(H_0+V_1)}-e^{-t(H_0+V_2)}\right\|_{\cJ_1}}\\
&\leq& \frac{t}{2}\int_0^1 d\tau\ \left\|e^{-t\tau(H_0+V_1)/2}\right\|_{2,2}\
\left\|e^{-t(1-\tau)(H_0+V_2)/2}\right\|_{2,2}\cdot\\
&&\cdot \left(\left\|e^{-t(H_0+V_1)/2}(V_1-V_2) \right\|_{\cJ_1} +
\left\|(V_1-V_2)e^{-t(H_0+V_2)/2} \right\|_{\cJ_1}
\right)\\
&\leq&
\frac{t}{2}\sup_{\tau\in(0,t)}\left\|e^{-\tau(H_0+V_1)/2}\right\|_{2,2}\
\sup_{\tau\in(0,t)}\left\|e^{-\tau(H_0+V_2)/2}\right\|_{2,2}\cdot\\
&&\cdot \left(\left\|e^{-t(H_0+V_1)/2}(V_1-V_2) \right\|_{\cJ_1} +
\left\|(V_1-V_2)e^{-t(H_0+V_2)/2} \right\|_{\cJ_1}
\right).
\end{eqnarray*}

Now we prove that for any $g\in l^1(L^2)$ and any $t>0$
\begin{displaymath}
\|g e^{-t(H_0+V)}\|_{\cJ_1}\leq c_t \left\|e^{-t(H_0+2V)}\right\|_{1,\infty}\|g\|_{l^1(L^2)}
\end{displaymath}
with a constant $c_t$ depending on $t$  only. We write
\begin{eqnarray*}
\lefteqn{g e^{-t(H_0+V)} = \sum_{\mathbf{j}\in\Z^\nu} g \chi_{\Delta_{\mathbf{j}}}
e^{-t(H_0+V)}} \\ &=&
\sum_{\mathbf{j}\in\Z^\nu} g \chi_{\Delta_{\mathbf{j}}} e^{-t(H_0+V)/2} (1+(\cdot-\mathbf{j})^2)^\nu\cdot
(1+(\cdot-\mathbf{j})^2)^{-\nu} e^{-t(H_0+V)/2},
\end{eqnarray*}
giving the a priori estimate
\begin{displaymath}
\|g e^{-t(H_0+V)}\|_{\cJ_1}\leq \sum_{\mathbf{j}\in\Z^\nu}
\left\|g\chi_{\Delta_{\mathbf{j}}}e^{-t(H_0+V)/2}(1+(\cdot-\mathbf{j})^2)^\nu\right\|_{\cJ_2}\
\left\|(1+(\cdot-\mathbf{j})^2)^{-\nu}e^{-t(H_0+V)/2}\right\|_{\cJ_2}.
\end{displaymath}
From the inequality \cite{Simon:79b,Simon:82}
\begin{equation}\label{lat.3.1}
0\leq e^{-t(H_0+V)}(x,y) \leq \left[e^{-t(H_0+2V)}(x,y)
\right]^{1/2}\left[e^{-tH_0}(x,y)\right]^{1/2},
\end{equation}
which is an easy consequence of the Feynman-Kac formula, we obtain
\begin{eqnarray}\label{lat.4}
e^{-t(H_0+V)}(x,y) &\leq& \left[\sup_{x,y\in\R^\nu}
e^{-t(H_0+2V)}(x,y)\right]^{1/2}\left[e^{-tH_0}(x,y)\right]^{1/2}\nonumber\\
&=&
\left\|e^{-t(H_0+2V)}
\right\|^{1/2}_{1,\infty}\left[e^{-tH_0}(x,y)\right]^{1/2},
\end{eqnarray}
and thus for any $h\in L^2$ we obtain
\begin{eqnarray*}
\left\|he^{-t(H_0+V)/2}\right\|_{\cJ_2} &\leq&\left\|e^{-t(H_0+2V)}
\right\|^{1/2}_{1,\infty}\ \left[\int_{\R^\nu}dx \int_{\R^\nu}dy\ |h(x)|^2
e^{-tH_0/2}(x,y)\right]^{1/2}\\ &\leq&
\left\|e^{-t(H_0+2V)}
\right\|^{1/2}_{1,\infty}\ \|h\|_{L^2}\ \left[\sup_x\int_{\R^\nu} dy\
e^{-tH_0/2}(x,y) \right]^{1/2}\\ &=&
\left\|e^{-t(H_0+2V)}
\right\|^{1/2}_{1,\infty}\ \|h\|_{L^2}\ \left\|e^{-tH_0/2} \right\|^{1/2}_{\infty,\infty}.
\end{eqnarray*}
Taking $h=(1+(\cdot-\mathbf{j})^2)^{-\nu}\in L^2(\R^\nu)$ we obtain
\begin{eqnarray*}
\lefteqn{\left\|(1+(\cdot-\mathbf{j})^2)^{-\nu} e^{-t(H_0+V)/2}\right\|_{\cJ_2}=
\left\|(1+(\cdot)^2)^{-\nu} e^{-t(H_0+V(\cdot-\mathbf{j})/2}\right\|_{\cJ_2}}\\
&\leq& \left\|e^{-t(H_0+2V)}
\right\|^{1/2}_{1,\infty}\ \left\|e^{-tH_0/2} \right\|^{1/2}_{\infty,\infty}\
\left(\int_{\R^\nu}\frac{dx}{(1+x^2)^{2\nu}} \right)^{1/2}.
\end{eqnarray*}
Now consider the operator $g\chi_{\Delta_{\mathbf{j}}}
e^{-t(H_0+V)/2}(1+(\cdot-\mathbf{j})^2)^\nu$ with an arbitrary $g\in l^1(L^2)$.
One has
\begin{eqnarray*}
\lefteqn{\left\|g\chi_{\Delta_{\mathbf{j}}}e^{-t(H_0+V)/2}(1+(\cdot-\mathbf{j})^2)^\nu\right\|_{\cJ_2}}\\
&\leq&\left\|g\chi_{\Delta_\mathbf{j}}(1+(\cdot-\mathbf{j})^2)^\nu\right\|_{L^2}\
\left\|\chi_{\Delta_{\mathbf{j}}}(1+(\cdot-\mathbf{j})^2)^{-\nu}e^{-t(H_0+V)/2}(1+(\cdot-\mathbf{j})^2)^\nu\right\|_{\cJ_2}\\
&\leq& \left(1+\frac{\nu}{4}\right)^\nu \|g\chi_{\Delta_{\mathbf{j}}}\|_{L^2}
\left\|\chi_{\Delta_{\mathbf{j}}}(1+(\cdot-\mathbf{j})^2)^{-\nu}e^{-t(H_0+V)/2}(1+(\cdot-\mathbf{j})^2)^\nu\right\|_{\cJ_2}.
\end{eqnarray*}
From the inequality (\ref{lat.4}) it follows that
\begin{eqnarray*}
 \lefteqn{(1+(x-\mathbf{j})^2)^{-\nu}\ e^{-t(H_0+V)/2}(x,y)\
(1+(y-\mathbf{j})^2)^{\nu}}\\ &\leq&
\left\|e^{-t(H_0+2V)/2}\right\|_{1,\infty}^{1/2}(1+(x-\mathbf{j})^2)^{-\nu}
\left[e^{-tH_0}(x,y) \right]^{1/2}(1+(y-\mathbf{j})^2)^{\nu}.
\end{eqnarray*}
Since $e^{-tH_0}(x,y)$ is translation invariant it suffices to estimate the
Hilbert-Schmidt norm of the integral operator with kernel $\chi_{\Delta_0}(x)
(1+x^2)^{-\nu} e^{-tH_0}(x,y) (1+y^2)^\nu$. From the inequality
\begin{displaymath}
(1+y^2)^\nu\leq C [(1+x^2)^\nu + |x-y|^{2\nu}]
\end{displaymath}
(see the proof of Lemma B.6.1 in \cite{Simon:82}) we obtain
\begin{eqnarray*}
\lefteqn{\chi_{\Delta_{\mathbf{0}}}(x)(1+x^2)^{-\nu}\left[e^{-tH_0}(x,y)\right]^{1/2}(1+y^2)^{\nu}}\\
&&\leq C\chi_{\Delta_{\mathbf{0}}}(x)\left[e^{-tH_0}(x,y)\right]^{1/2} + C
\chi_{\Delta_{\mathbf{0}}}(x)(1+x^2)^{-\nu}
|x-y|^{2\nu}\left[e^{-tH_0}(x,y)\right]^{1/2},
\end{eqnarray*}
which is obviously square integrable with respect to the measure $dx dy$.
\end{proof}

We will need a weaker form of (\ref{lat.3}). First we note that by the
semigroup property and by the duality
($\|e^{-tH}\|_{1,2}=\|e^{-tH}\|_{2,\infty}$ since $e^{-tH}$ is self-adjoint) we
have
\begin{displaymath}
\left\|e^{-t(H_0+V)} \right\|_{1,\infty}\leq \left\|e^{-t(H_0+V)/2} \right\|_{1,2}
\left\|e^{-t(H_0+V)/2} \right\|_{2,\infty}=\left\|e^{-t(H_0+V)/2} \right\|_{2,\infty}^2.
\end{displaymath}
By Lemma \ref{A.3:neu}
\begin{displaymath}
\left\|e^{-t(H_0+V)} \right\|_{2,\infty}^2\leq (4\pi t)^{-\nu/2}
\left\|e^{-t(H_0+2V)} \right\|_{\infty,\infty}.
\end{displaymath}
Since $\|e^{-t(H_0+V)}\|_{2,2}\leq \|e^{-t(H_0+V)}\|_{\infty,\infty}$ (see
Theorem \ref{A.2:neu}) from Lemma \ref{th.SSD.4} it follows that
\begin{eqnarray}\label{lat.5}
\lefteqn{\left\|e^{-t(H_0+V_1)}-e^{-t(H_0+V_2)}\right\|_{\cJ_1}}\nonumber\\ &\leq&
C_t (2\pi t)^{-\nu}
\sup_{\tau\in(0,t)}\left\|e^{-\tau(H_0+V_1)/2}\right\|_{\infty,\infty}\
\sup_{\tau\in(0,t)}\left\|e^{-\tau(H_0+V_2)/2}\right\|_{\infty,\infty}\cdot\nonumber\\
&&\cdot\left\|e^{-t(H_0+4V_1)/4}\right\|_{\infty,\infty}\
\left\|e^{-t(H_0+4V_2)/4}\right\|_{\infty,\infty}\
\|V_1-V_2\|_{l^1(L^2)}.
\end{eqnarray}
By the inequality (\ref{SSG3}) both suprema are finite.

\begin{lemma}\label{th.SSD.5}
Let $f\in L^1(\R^\nu)$. For any sequence of boxes $\Lambda$ such that
$\Lambda\rightarrow\infty$
\begin{displaymath}
\lim_{\Lambda\rightarrow\infty}\frac{1}{\meas(\Lambda)}\int_{\Lambda}dx \int_{\Lambda^c}dy f(x-y)=0.
\end{displaymath}
A similar statement holds in the discrete case. If $f\in l^1(\Z^\nu)$ then
\begin{displaymath}
\lim_{\Lambda\rightarrow\infty}\frac{1}{\#\{\mathbf{j}\in \Lambda\}}\sum_{\substack{
\mathbf{j}\in\Z^\nu \\ \mathbf{j}\in \Lambda}}
\sum_{\substack{\mathbf{k}\in\Z^\nu \\ \mathbf{k}\notin \Lambda}} f(\mathbf{j}-\mathbf{k}) = 0.
\end{displaymath}
\end{lemma}

Certainly this lemma remains valid for much more general domains  than boxes,
but we will not go in the details here.

\begin{remark}\label{rem.SSD.6}
Let $\nu\geq 2$. Suppose that $f$ is integrable with an exponential weight,
$f\in L^1(\R^\nu; e^{\alpha|x|}dx)$. Then
\begin{displaymath}
\int_{\Lambda}dx \int_{\Lambda^c}dy f(x-y)= O\left(\meas_{\nu-1}(\partial \Lambda)\right).
\end{displaymath}
In the discrete case $f\in l^1(\Z^\nu; e^{\alpha|\mathbf{j}|})$ implies that
\begin{displaymath}
\sum_{\substack{\mathbf{j}\in\Z^\nu \\ \mathbf{j}\in \Lambda}}
\sum_{\substack{\mathbf{k}\in\Z^\nu \\ \mathbf{k}\notin \Lambda}}
f(\mathbf{j}-\mathbf{k}) = O\left(\meas_{\nu-1}(\partial \Lambda)\right).
\end{displaymath}
\end{remark}

\begin{proof}[Proof of Lemma \ref{th.SSD.5}]
Without loss of generality we may suppose that $f\geq 0$. First we consider the
case $\nu=1$. It suffices to prove that
\begin{equation}\label{to.prove}
\lim_{R\rightarrow\infty}\frac{1}{R}\int_0^R dx \int_R^\infty dy f(x-y)=0.
\end{equation}
Obviously
\begin{displaymath}
\int_0^R dx \int_R^\infty f(x-y) dy = \int_0^R F(-x)dx= R\int_0^1 F(-xR)dx,
\end{displaymath}
where
\begin{displaymath}
F(x)=\int_{-\infty}^x f(y)dy.
\end{displaymath}
The function $F(x)$ is monotone non-decreasing, $F(-\infty)=0$, and
$F(\infty)<\infty$.  Therefore $F(-xR)\leq F(-x)$ for all $x\in[0,1]$ and
$R\geq 1$. Since $F(-xR)\rightarrow 0$ pointwise as $R\rightarrow\infty$ by the
Lebesgue dominated convergence theorem we obtain
\eqref{to.prove}.

Now we turn to the case $\nu\geq 2$. According to the decomposition
$\R^\nu=\R\oplus\R^{\nu-1}$ we represent $\Lambda=\Lambda_1\times\Lambda_2$.
Obviously,
\begin{eqnarray*}
\int_\Lambda dx \int_{\Lambda^c} dy f(x-y) &\leq& \int_{\Lambda_1}dx_1 \int_{\Lambda_2}dx_2
\int_{\Lambda_1^c}dy_1 \int_{\R^{\nu-1}}dy_2 f(x-y)\\
&=& \meas_{\nu-1}(\Lambda_2) \int_{\Lambda_1}dx_1 \int_{\Lambda_1^c}dy_1
\widetilde{f}(x_1-y_1),
\end{eqnarray*}
where
\begin{displaymath}
\widetilde{f}(x_1-y_1)=\int_{\R^{\nu-1}} dy_2 f(x-y).
\end{displaymath}
By the Fubini theorem $\widetilde{f}\in L^1(\R)$. Since
$\meas_\nu(\Lambda)=\meas_1(\Lambda_1)\ \meas_{\nu-1}(\Lambda_2)$ by
\eqref{to.prove} the claim follows. In the discrete case the claim can be proved
in the same way.
\end{proof}

\begin{proof}[Proof of Theorem \ref{th.SSD:main}]
For simplicity we consider the case $\L^\nu=\Z^\nu$. The general case can be
considered in the same way. In the estimate \eqref{lat.5} we set
$V_1=\chi_\Lambda V$ and $V_2=V_\Lambda$. By the monotonicity property of the
Schr\"{o}dinger semigroups \eqref{SSG2} we have
\begin{eqnarray*}
\left\|e^{-t(H_0+\chi_\Lambda V)}\right\|_{\infty,\infty} &\leq&
\left\|e^{-t(H_0+V_-)}\right\|_{\infty,\infty},\\
\left\|e^{-t(H_0+V_\Lambda)}\right\|_{\infty,\infty} &\leq&
\left\|e^{-t(H_0+V_-)}\right\|_{\infty,\infty}
\end{eqnarray*}
for all $\Lambda$'s. Since $V_-\in K_\nu$ the norm
$\|e^{-t(H_0+V_-)}\|_{\infty,\infty}$ is finite for all $t>0$. Thus it follows
that for any $t>0$ there is a constant $C>0$ independent of $\Lambda$ such that
\begin{displaymath}
\left\|e^{-t(H_0+\chi_\Lambda V)-e^{-t(H_0+V_\Lambda)}}\right\|_{\cJ_1}
\leq C \|\chi_\Lambda V- V_\Lambda\|_{l^1(L^2)}.
\end{displaymath}
Obviously we have
\begin{eqnarray*}
\|\chi_\Lambda V- V_\Lambda\|_{l^1(L^2)} &\leq& \Big\|(1-\chi_\Lambda)
\sum_{\substack{\mathbf{j}\in\Z^\nu \\ \mathbf{j}\in\Lambda}}
f_{\mathbf{j}}(\cdot-\mathbf{j})\Big\|_{l^1(L^2)} + \Big\|\chi_\Lambda
\sum_{\substack{\mathbf{j}\in\Z^\nu \\ \mathbf{j}\in\Lambda^c}}
f_{\mathbf{j}}(\cdot-\mathbf{j})\Big\|_{l^1(L^2)}\\ &\leq &
\sum_{\substack{\mathbf{j}\in\Z^\nu \\ \mathbf{j}\in\Lambda}}
\|(1-\chi_\Lambda)f_{\mathbf{j}}(\cdot-\mathbf{j})\|_{l^1(L^2)}+
\sum_{\substack{\mathbf{j}\in\Z^\nu \\ \mathbf{j}\in\Lambda^c}}
\|\chi_\Lambda f_{\mathbf{j}}(\cdot-\mathbf{j})\|_{l^1(L^2)}.
\end{eqnarray*}
Without loss of generality we can choose boxes $\Lambda$ such that
\begin{displaymath}
1-\chi_\Lambda = \sum_{\substack{\mathbf{j}\in\Z^\nu \\ \mathbf{j}\in
\Lambda^c}} \chi_{\Delta_{\mathbf{j}}} \qquad \mathrm{and} \qquad
\chi_\Lambda = \sum_{\substack{\mathbf{j}\in\Z^\nu \\ \mathbf{j}\in
\Lambda}} \chi_{\Delta_{\mathbf{j}}}
\end{displaymath}
and then we obtain that the r.h.s.\ of this inequality is bounded by
\begin{eqnarray}\label{lat.6}
&&\sum_{\substack{\mathbf{j}\in\Z^\nu \\ \mathbf{j}\in\Lambda}}
\sum_{\substack{\mathbf{k}\in\Z^\nu \\ \mathbf{k}\in\Lambda^c}}
\|\chi_{\Delta_{\mathbf{k}}} f_{\mathbf{j}}(\cdot-\mathbf{j})\|_{l^1(L^2)}
+\sum_{\substack{\mathbf{j}\in\Z^\nu \\ \mathbf{j}\in\Lambda^c}}
\sum_{\substack{\mathbf{k}\in\Z^\nu \\ \mathbf{k}\in\Lambda}}
\|\chi_{\Delta_{\mathbf{k}}} f_{\mathbf{j}}(\cdot-\mathbf{j})\|_{l^1(L^2)}
\nonumber\\
&=&\sum_{\substack{\mathbf{j}\in\Z^\nu \\ \mathbf{j}\in\Lambda}}
\sum_{\substack{\mathbf{k}\in\Z^\nu \\ \mathbf{k}\in\Lambda^c}}
\|\chi_{\Delta_{\mathbf{k}}} f_{\mathbf{j}}(\cdot-\mathbf{j})\|_{L^2}
+\sum_{\substack{\mathbf{j}\in\Z^\nu \\ \mathbf{j}\in\Lambda^c}}
\sum_{\substack{\mathbf{k}\in\Z^\nu \\ \mathbf{k}\in\Lambda}}
\|\chi_{\Delta_{\mathbf{k}}} f_{\mathbf{j}}(\cdot-\mathbf{j})\|_{L^2}
\nonumber\\
&=&\sum_{\substack{\mathbf{j}\in\Z^\nu \\ \mathbf{j}\in\Lambda}}
\sum_{\substack{\mathbf{k}\in\Z^\nu \\ \mathbf{k}\in\Lambda^c}}
\|\chi_{\Delta_{\mathbf{k}-\mathbf{j}}} f_{\mathbf{j}}\|_{L^2}+
\sum_{\substack{\mathbf{j}\in\Z^\nu \\ \mathbf{j}\in\Lambda^c}}
\sum_{\substack{\mathbf{k}\in\Z^\nu \\ \mathbf{k}\in\Lambda}}
\|\chi_{\Delta_{\mathbf{k}-\mathbf{j}}} f_{\mathbf{j}}\|_{L^2},
\end{eqnarray}
where in the last step we have used the invariance of the norm with respect to
translations and the fact that $\chi_{\Delta_{\mathbf{k}}}(x+\mathbf{j})=
\chi_{\Delta_{\mathbf{k}-\mathbf{j}}}(x)$. The assumption that the family $f_{\mathbf{j}}$
is uniformly in $l^1(L^2)$ (see \eqref{lat.1}) implies that
\begin{displaymath}
g_{\mathbf{j}}=\sup_{\mathbf{k}\in\Z^\nu}\|\chi_{\Delta_{\mathbf{j}}}f_{\mathbf{k}}\|_{L^2},
\quad \mathbf{j}\in\Z^\nu
\end{displaymath}
is summable, i.e.\ $g\in l^1(\Z^\nu)$. Since
$\|\chi_{\Delta_{\mathbf{k}-\mathbf{j}}}f_{\mathbf{j}}\|_{L^2}\leq
g_{\mathbf{k}-\mathbf{j}}$ we can estimate the r.h.s.\ of \eqref{lat.6} by
\begin{displaymath}
\sum_{\substack{\mathbf{j}\in\Z^\nu \\ \mathbf{j}\in\Lambda}}
\sum_{\substack{\mathbf{k}\in\Z^\nu \\ \mathbf{k}\in\Lambda^c}} g_{\mathbf{k}-\mathbf{j}}
+\sum_{\substack{\mathbf{j}\in\Z^\nu \\ \mathbf{j}\in\Lambda^c}}
\sum_{\substack{\mathbf{k}\in\Z^\nu \\ \mathbf{k}\in\Lambda}} g_{\mathbf{k}-\mathbf{j}}.
\end{displaymath}
Applying now Lemma \ref{th.SSD.5} we obtain
\begin{displaymath}
\lim_{\Lambda\rightarrow\infty} (\meas(\Lambda))^{-1}
\tr\left[e^{-t(H_0+\chi_\Lambda V)-e^{-t(H_0+V_\Lambda)}} \right]=0.
\end{displaymath}
Now applying the arguments used to prove Theorem \ref{th.SSD.1} completes the
proof.
\end{proof}

%%%%%%%%%%%%%%%%%%%%%%%%%%%%%%%%%%%%%%%%%%%%%%%%%%%%%%
\section{Cluster Properties of the Spectral Shift Function}
%%%%%%%%%%%%%%%%%%%%%%%%%%%%%%%%%%%%%%%%%%%%%%%%%%%%%%
\setcounter{equation}{0}

Consider a potential $V$ different from zero on a set of positive Lebesgue
measure such that $V_-\in K_\nu$ and $V_+\in K_\nu^{\mathrm{loc}}$. Let
$\Lambda$ be an arbitrary open set such that $\Int(\supp V)\subseteq \Lambda$.
Consider some decomposition of $\Lambda$ into two disjoint parts $\Lambda_1$
and $\Lambda_2$ such that $\Lambda=\Int(\overline{\Lambda_1\cup\Lambda_2})$.

\begin{definition}\label{def:extension}
We call the open sets $\widetilde{\Lambda_1}$ and $\widetilde{\Lambda_2}$
\textbf{complete extensions} of $\Lambda_1$ and $\Lambda_2$ respectively iff

\noindent (i) \ \ \ $\overline{\widetilde{\Lambda_1}\cup\widetilde{\Lambda_2}}=\R^\nu$,

\noindent (ii) \ \ $\Lambda_1\subseteq\widetilde{\Lambda_1}$, \qquad\qquad\qquad\ \
$\Lambda_2\subseteq\widetilde{\Lambda_2}$

\noindent (iii) \ $\overline{\widetilde{\Lambda_1}}\cap\overline{\Lambda_2} =
\overline{\Lambda_1}\cap\overline{\Lambda_2}$, \qquad
$\overline{\widetilde{\Lambda_2}}\cap\overline{\Lambda_1} =
\overline{\Lambda_1}\cap\overline{\Lambda_2}$.
\end{definition}

\begin{remark}\label{rem:extension}
The condition (iii) says that the common boundary of $\overline{\Lambda_1}$ and
$\overline{\Lambda_1}$ is the same as that of
$\overline{\widetilde{\Lambda_1}}$ and $\overline{\Lambda_2}$ and of
$\overline{\widetilde{\Lambda_2}}$ and $\overline{\Lambda_1}$.
\end{remark}

\begin{example}\label{ex:extension}
\begin{figure}[htb]
\centerline{\epsfig{figure=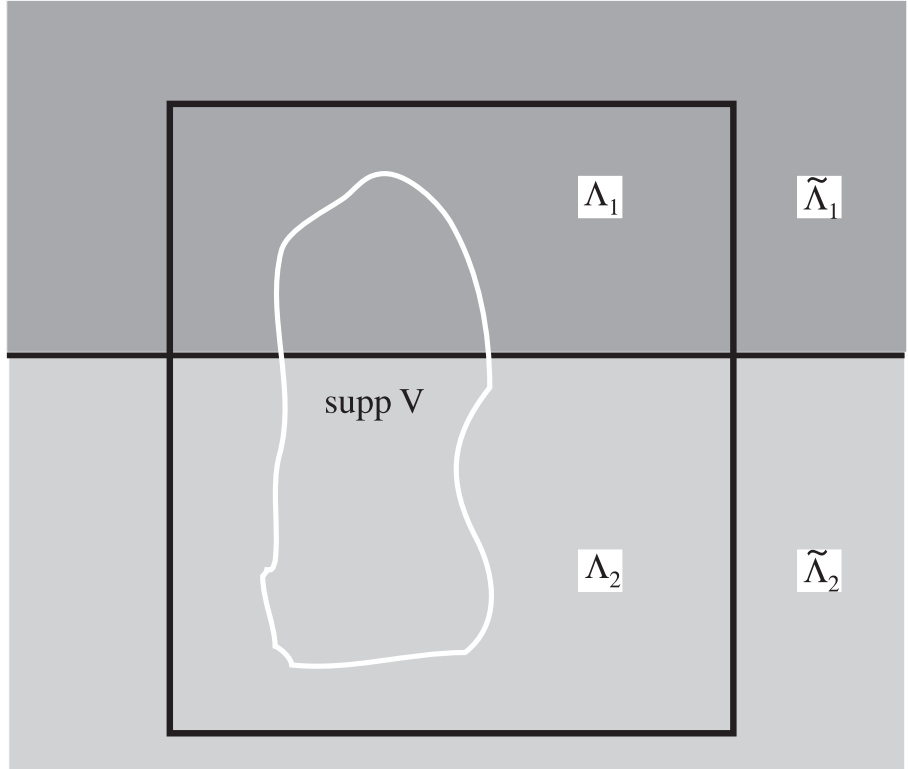,width=120mm,clip=}}
\caption{\label{fig:extension} Illustration to the Example \ref{ex:extension}.}
\end{figure}
Consider some $V$ with compact support and choose a box $\Lambda$ such that
$\supp V\subset \Lambda$. Take an arbitrary hyperplane dividing $\Lambda$ into
two parts, the interiors of which we denote by $\Lambda_1$ and $\Lambda_2$.
Complete extensions $\widetilde{\Lambda_1}$ and $\widetilde{\Lambda_2}$ are
simply the open half-spaces containing $\Lambda_1$ and $\Lambda_2$ respectively
(see Fig.\ \ref{fig:extension}).
\end{example}

\begin{theorem}\label{th:extension}
Let $V$ be a potential with compact support such that $V_+\in
K_\nu^{\mathrm{loc}}$ and $V_-\in K_\nu$. For any $t>0$ and arbitrary domains
$\Lambda_1,\Lambda\subset\R^\nu$ such that
$\overline{\Lambda_1\cup\Lambda_2}\supseteq\supp V$
\begin{eqnarray*}
\lefteqn{\left\|e^{-t(H_0+V)}-e^{-t(H_0+\chi_{\Lambda_1}V)}-e^{-t(H_0+\chi_{\Lambda_2}V)}
+e^{-tH_0} \right\|_{\cJ_1}}\\ &\leq& 2^{3-\nu/4}(\pi t)^{-\nu/2}
\left\|e^{-t(H_0+2V_-)/2}
\right\|_{\infty,\infty}^{1/2}\ \left\|e^{-t(H_0+4V_-)/4}
\right\|_{\infty,\infty}^{1/2}\\
&&\cdot\big(\|\chi_{\widetilde{\Lambda_1}}
\mathbf{P}_\bullet\{\tau_{\Lambda_2}\leq t/2\}^{1/2}\|_{L^1}+
\|\mathbf{E}_\bullet\{\chi_{\widetilde{\Lambda_1}}; \tau_{\Lambda_2}\leq t/2\}^{1/2}\|_{L^1}\\
&& + \|\chi_{\widetilde{\Lambda_2}}
\mathbf{P}_\bullet\{\tau_{\Lambda_1}\leq t/2\}^{1/2}\|_{L^1}+
\|\mathbf{E}_\bullet\{\chi_{\widetilde{\Lambda_2}}; \tau_{\Lambda_1}\leq t/2\}^{1/2}\|_{L^1}\big),
\end{eqnarray*}
where $\widetilde{\Lambda_1}$ and $\widetilde{\Lambda_2}$ are complete
extensions of $\Lambda_1$ and $\Lambda_2$ respectively.
\end{theorem}

\begin{proof}
We write $V_i=\chi_{\Lambda_i}V$, $i=1,2$ such that $V=V_1+V_2$ and
\begin{eqnarray}\label{extension.1}
\lefteqn{e^{-t(H_0+V_1+V_2)}-e^{-t(H_0+V_1)}-e^{-t(H_0+V_2)}+e^{-tH_0}}\nonumber\\
&=& \chi_{\widetilde{\Lambda_1}}\left(e^{-t(H_0+V_1+V_2)}-e^{-t(H_0+V_1)}
\right)
 +\chi_{\widetilde{\Lambda_2}}\left(e^{-t(H_0+V_1+V_2)}-e^{-t(H_0+V_2)}
\right)\nonumber\\
&& - \chi_{\widetilde{\Lambda_1}}\left(e^{-t(H_0+V_2)}-e^{-tH_0} \right)
- \chi_{\widetilde{\Lambda_2}}\left(e^{-t(H_0+V_1)}-e^{-tH_0} \right).
\end{eqnarray}
Consider the first term on the r.h.s.\ of this expression. We represent it in
the form
\begin{eqnarray*}
\chi_{\widetilde{\Lambda_1}}\left(e^{-t(H_0+V_1+V_2)}-
e^{-t(H_0+V_1+\infty_{\Lambda_2})} \right)
-\chi_{\widetilde{\Lambda_1}}\left(e^{-t(H_0+V_1)}-
e^{-t(H_0+V_1+\infty_{\Lambda_2})}\right).
\end{eqnarray*}
The proof now closely follows along the lines of the proof of Lemma
\ref{thm:Stollmann:5}. Denoting
\begin{displaymath}
D(t)=e^{-t(H_0+V_1+V_2)}-e^{-t(H_0+V_1+\infty_{\Lambda_2})}
\end{displaymath}
we obtain
\begin{eqnarray*}
\|\chi_{\widetilde{\Lambda_1}}D(t)\|_{\cJ_1} &\leq & \|D(t/2)^2
\chi_{\widetilde{\Lambda_1}}\|_{\cJ_1}\\
&& + \left\| e^{-t(H_0+V_1+\infty_{\Lambda_2})}\chi_{\widetilde{\Lambda_1}}
D(t/2)\chi_{\widetilde{\Lambda_1}}\right\|_{\cJ_1}\\ &&
+\left\|e^{-t(H_0+V_1+\infty_{\Lambda_2})}\chi_{\widetilde{\Lambda_1}}
D(t/2)\right\|_{\cJ_1}.
\end{eqnarray*}
For an arbitrary $f\in L^2(\R^\nu)$ with $\|f\|_{L^2}\leq 1$ we have
\begin{displaymath}
|(D(t)f)(x)|\leq \left\|e^{-t(H_0+2V_-)} \right\|_{1,\infty}^{1/2}
(\mathbf{P}_x\{\tau_{\Lambda_2}\leq t\})^{1/2}
\end{displaymath}
and analogously
\begin{displaymath}
|(D(t)\chi_{\widetilde{\Lambda_1}}f)(x)|\leq \left\|e^{-t(H_0+2V_-)}
\right\|_{1,\infty}^{1/2}
(\mathbf{E}_x\{\chi_{\widetilde{\Lambda_1}}(X_t); \tau_{\Lambda_2}\leq
t\})^{1/2}.
\end{displaymath}
Now by Lemma \ref{Stollmann:Tr} it follows that
\begin{eqnarray*}
\lefteqn{\left\|\chi_{\widetilde{\Lambda_1}}\left(e^{-t(H_0+V_1+V_2)}-
e^{-t(H_0+V_1+\infty_{\Lambda_2})}\right) \right\|_{\cJ_1}}\\ &\leq& 2
\left\|e^{-t(H_0+V_-)/2}\right\|_{1,2}\ \left\|e^{-t(H_0+2V_-)/2}\right\|_{1,\infty}^{1/2}\\
&&
\cdot\left(\|\chi_{\widetilde{\Lambda_1}}\mathbf{P}_\bullet\{\tau_{\Lambda_2}\leq t/2
\}^{1/2}\|_{L^1}+\|\mathbf{E}_\bullet\{\chi_{\widetilde{\Lambda_1}}(X_t);
\tau_{\Lambda_2}\leq t/2 \}^{1/2}\|_{L^1}
\right).
\end{eqnarray*}
Similarly we obtain
\begin{eqnarray*}
\lefteqn{\left\|\chi_{\widetilde{\Lambda_1}}\left(e^{-t(H_0+V_1)}-
e^{-t(H_0+V_1+\infty_{\Lambda_2})}\right) \right\|_{\cJ_1}}\\ &\leq& 2
\left\|e^{-t(H_0+V_-)/2}\right\|_{1,2}\ \left\|e^{-t(H_0+2V_-)/2}\right\|_{1,\infty}^{1/2}\\
&&
\cdot\left(\|\chi_{\widetilde{\Lambda_1}}\mathbf{P}_\bullet\{\tau_{\Lambda_2}\leq t/2
\}^{1/2}\|_{L^1}+\|\mathbf{E}_\bullet\{\chi_{\widetilde{\Lambda_1}}(X_t);
\tau_{\Lambda_2}\leq t/2 \}^{1/2}\|_{L^1}
\right).
\end{eqnarray*}
Finally as in the proof of Lemma \ref{thm:Stollmann:5} we obtain
\begin{eqnarray*}
\|\chi_{\widetilde{\Lambda_1}}D(t)\|_{\cJ_1} \leq 2^{1-\nu/4}(\pi t)^{-\nu/2}
\left\|e^{-t(H_0+2V_-)/2}\right\|_{\infty,\infty}^{1/2}\
\left\|e^{-t(H_0+4V_-)/4}\right\|_{\infty,\infty}^{1/2}\\
\cdot \left(\|\chi_{\widetilde{\Lambda_1}}\mathbf{P}_\bullet\{\tau_{\Lambda_2}\leq t/2
\}^{1/2}\|_{L^1}+\|\mathbf{E}_\bullet\{\chi_{\widetilde{\Lambda_1}}(X_t);
\tau_{\Lambda_2}\leq t/2 \}^{1/2}\|_{L^1}
\right).
\end{eqnarray*}
The other terms on the r.h.s.\ of (\ref{extension.1}) can be estimated in a
similar way.
\end{proof}

Due to Lemmas \ref{thm:Stollmann:2} and \ref{thm:Stollmann:3.1} from Theorem
\ref{th:extension} follows

\begin{corollary}\label{cor:extension}
Let $\Lambda$, $\Lambda_1$ and $\Lambda_2$ be such that as in Example
\ref{ex:extension}. If $\nu\geq 2$ then for any $t>0$ there is a constant $c>0$
depending on $t$ only such that
\begin{displaymath}
\left\|e^{-t(H_0+V)}-e^{-t(H_0+\chi_{\Lambda_1}V)}-
e^{-t(H_0+\chi_{\Lambda_2}V)}+e^{-tH_0}\right\|_{\cJ_1}\leq c\
\meas_{\nu-1}(\overline{\Lambda_1}\cap\overline{\Lambda_2}).
\end{displaymath}
If $\nu=1$ the same inequality holds if its r.h.s.\ is replaced by some
constant.
\end{corollary}

Corollary \ref{cor:extension} implies that for every $t>0$
\begin{eqnarray*}
\left|\int_\R e^{-t\lambda}\left(\xi(\lambda;H_0+V,H_0)-
\xi(\lambda;H_0+\chi_{\Lambda_1}V, H_0) -
\xi(\lambda;H_0+\chi_{\Lambda_2}V, H_0)\right)d\lambda \right| \\ \leq c\
\meas_{\nu-1}(\overline{\Lambda_1}\cap\overline{\Lambda_2}).
\end{eqnarray*}
It is natural to pose the question whether such estimates also hold in the
pointwise sense (i.e. for the spectral shift functions itself). The following
example shows that the answer is in general negative.

\begin{example}\label{Kirsch}
Consider the hypercube $C_L$ in $\R^\nu$, $\nu\geq 2$ centered at the origin
with side length $L$. Denote by $H_{0L}$ minus the Laplacian on $C_L$ with
Dirichlet boundary conditions on $\partial C_L$, i.e.
$H_{0L}=-\Delta+\infty_{C_L^c}$. Let $V$ be a bounded non-negative potential
with support in the unit cube centered at the origin. Let $E_n(H)$,
$n=0,1,\ldots$ be the eigenvalues of a semibounded from below operator $H$
counted in increasing order taking into account their multiplicities. Let
$N(\lambda;H)=\#\{n|\ E_n(H)\leq \lambda\}$ be the corresponding counting
function. Kirsch \cite{Kirsch:87} proved that the difference
\begin{displaymath}
\phi_L(\lambda)=N(\lambda;H_{0L})-N(\lambda;H_{0L}+V)\geq 0
\end{displaymath}
is an unbounded function with respect to $L>1$ for any $\lambda>0$, i.e.\
\begin{displaymath}
\sup_{L>1}\phi_L(\lambda)=\infty.
\end{displaymath}
This obviously implies that the difference of the spectral shift functions
\begin{eqnarray*}
\lefteqn{\psi_L(\lambda)=\xi(\lambda;H_{0L}+V,H_{0L})-\xi(\lambda;H_0+V,H_0)}\\
&=&\xi(\lambda;H_{0L}+V,H_{0})-\xi(\lambda;H_{0L},H_0)-\xi(\lambda;H_0+V,H_0)
\end{eqnarray*}
is unbounded with respect to $L>1$ for any $\lambda>0$. On the other hand using
the technique from the proof of Theorem \ref{th:extension} one can prove that
its Laplace transform
\begin{displaymath}
\Psi_L(t)=\int_0^\infty e^{-\lambda t} \psi_L(\lambda)d\lambda
\end{displaymath}
is uniformly bounded with respect to $L>1$ for every fixed $t>0$.
\end{example}

%%%%%%%%%%%%%%%%%%%%%%%%%%%%%%%%%%%%%%%%%%%%%%%%%%%%%%
\section{Applications to Random Schr\"{o}dinger Operators}
%%%%%%%%%%%%%%%%%%%%%%%%%%%%%%%%%%%%%%%%%%%%%%%%%%%%%%
\setcounter{equation}{0}

\subsection{Random Potential on Lattices}

Here we consider random potentials of the form
\begin{equation}\label{ran:pot:def}
V_\omega(x) = \sum_{\mathbf{j}\in \Z^\nu} \alpha_{\mathbf{j}}(\omega)
f(\cdot-\mathbf{j}),
\end{equation}
where $\alpha_{\mathbf{j}}(\omega)$ is a sequence of random i.i.d.\ variables
on a probability space ($\Omega,\mathfrak{F},\P$) with common distribution
$\kappa$, i.e.\ $\mathfrak{F}$ is a $\sigma$-algebra on $\Omega$, $\P$ a
probability measure on ($\Omega,\mathfrak{F}$) and
$\kappa(B)=\P\{\alpha_{\mathbf{j}}\in B\}$ for any Borel subset $B$ of $\R$.
Let $\E$ denote the expectation with respect to $\P$. The random variables
$\{\alpha_{\mathbf{j}}(\omega)\}_{\mathbf{j}\in\Z^\nu}$ are supposed to form a
stationary, metrically transitive random field, i.e.\ there are measure
preserving ergodic transformations $\{T_{\mathbf{j}}\}_{\mathbf{j}\in\Z^\nu}$
such that
$\alpha_{\mathbf{j}}(T_{\mathbf{k}}\omega)=\alpha_{\mathbf{j}-\mathbf{k}}(\omega)$
for all $\omega\in\Omega$. The single-site potential $f$ is supposed to be
supported in the unit cube $\Delta_0$ centered at the origin, $\supp f\subseteq
\Delta_0=[-1/2,1/2]^\nu$ and $f\in L^2(\R^\nu)$. Additionally if $\nu\geq 4$
the potential $f$ is supposed to belong to $L^p(\R^\nu)$ with some $p>\nu/2$.
Instead of the integer lattice in (\ref{ran:pot:def}) we can consider an
arbitrary lattice $\L^\nu$ as discussed in Section \ref{sec:lattices}.

Finally if $f$ is sign-indefinite, i.e.\ both $f>0$ and $f<0$ on sets of
positive Lebesgue measure, in this section we will suppose that $\supp\ \kappa$
is bounded, i.e.\ there are finite $\alpha_\pm$ such that
$\alpha_-\leq\alpha_{\mathbf{j}}(\omega)\leq\alpha_+$ for all
$\mathbf{j}\in\Z^\nu$ and all $\omega\in\Omega$. Also if $f\geq 0$ ($f\leq 0$)
then $\supp\ \kappa$ is supposed to be bounded below (above), i.e.\ there is
$\alpha_->-\infty$ ($\alpha_+<\infty$) such that
$\alpha_{\mathbf{j}}(\omega)\geq\alpha_-$
($\alpha_{\mathbf{j}}(\omega)\leq\alpha_+$) for all $\mathbf{j}\in\Z^\nu$ and
all $\omega\in\Omega$. These conditions can be relaxed by requiring that the
expectations of certain quantities are finite. The corresponding modifications
are obvious and we will not dwell on them.

For an arbitrary box $\Lambda$ we consider
\begin{equation}\label{ran:pot:def1}
V_{\omega,\Lambda}(x) = \sum_{\substack{\mathbf{j}\in \Z^\nu \\
\mathbf{j}\in\Lambda}}
\alpha_{\mathbf{j}}(\omega) f(\cdot-\mathbf{j}).
\end{equation}
For any $t>0$ denote
\begin{eqnarray*}
\cF_{\omega,\Lambda}(t)=\tr\left(e^{-t(H_0+V_{\omega,\Lambda})}-e^{-tH_0}\right)
=-t \int_\R e^{-\lambda t} \xi(\lambda; H_0+V_{\omega,\Lambda}, H_0)d\lambda.
\end{eqnarray*}

We note that for arbitrary translations $U(d)$, $d\in\R^\nu$,
$(U(d)f)(x)=f(x-d)$ one has
\begin{displaymath}
\tr\left(e^{-t(H_0+U^{-1}VU)}-e^{-tH_0} \right) = \tr\left(e^{-t(H_0+V)}-e^{-tH_0} \right).
\end{displaymath}
Thus the metrical transitivity of $\alpha_{\mathbf{j}}(\omega)$ implies that
\begin{equation}\label{metr:trans}
\cF_{T_{\mathbf{j}}\omega,\Lambda}(t)=\cF_{\omega,\Lambda-\mathbf{j}}(t).
\end{equation}

By the monotonicity property (\ref{SSG2})
$\sup_\Lambda\|e^{-t(H_0+V_\Lambda)}\|_{\infty,\infty}$ is finite. Therefore
from Corollary \ref{cor:extension} it follows that for any $t>0$ there is a
constant $C$ such that
\begin{equation}\label{ran:pot:2}
\left| \cF_{\omega,\Lambda}(t)- \cF_{\omega,\Lambda_1}(t)-\cF_{\omega,\Lambda_2}(t)\right|
\leq C \meas_{\nu-1}(S_{12})
\end{equation}
for any boxes $\Lambda_1$ and $\Lambda_2$ such that
$\Lambda_1\cup\Lambda_2=\Lambda$ and where $S_{12}$ denotes the common surface
of $\Lambda_1$ and $\Lambda_2$.

Let
\begin{displaymath}
\cF^{\pm}_{\omega,\Lambda}(t) = \cF_{\omega,\Lambda}(t) \pm \frac{C}{2}
\meas_{\nu-1}(\partial\Lambda).
\end{displaymath}
From the inequalities (\ref{ran:pot:2}) it follows that for every fixed $t>0$
$\cF^{+}(t)$ is subadditive whereas $\cF^{-}(t)$ is superadditive with respect
to $\Lambda$. Indeed, e.g.\ for $\cF^{+}(t)$ we have
\begin{eqnarray*}
\cF^{+}_{\omega,\Lambda}(t)-\cF^{+}_{\omega,\Lambda_1}(t)-\cF^{+}_{\omega,\Lambda_2}(t)
\qquad\qquad\qquad\qquad\qquad\qquad\qquad\qquad\qquad\qquad\\
=\cF_{\omega,\Lambda}(t)-\cF_{\omega,\Lambda_1}(t)-\cF_{\omega,\Lambda_2}(t)
+\frac{C}{2}\left(\meas_{\nu-1}(\partial\Lambda)-\meas_{\nu-1}(\partial\Lambda_1)
-\meas_{\nu-1}(\partial\Lambda_2)
\right)\\
\leq C\ \meas_{\nu-1}(S_{12})- C\ \meas_{\nu-1}(S_{12}) = 0.
\end{eqnarray*}

Now we show that
\begin{eqnarray*}
\Gamma_+ = \inf_\Lambda \frac{1}{\meas(\Lambda)}\E\{\cF_{\omega,\Lambda}^+(t)\} > -\infty
\end{eqnarray*}
and
\begin{eqnarray*}
\Gamma_- = \sup_\Lambda \frac{1}{\meas(\Lambda)}\E\{\cF_{\omega,\Lambda}^-(t)\} < \infty.
\end{eqnarray*}
To this end we note that
\begin{eqnarray*}
\Gamma_- &=& \sup_\Lambda \frac{1}{\meas(\Lambda)}\E\left\{\cF_{\omega,\Lambda}(t)
-\frac{C}{2} \meas_{\nu-1}(\partial\Lambda)\right\}\\
&\leq& \sup_\Lambda \frac{1}{\meas(\Lambda)}\E\{\cF_{\omega,\Lambda}^+(t)\}
\leq \sup_\Lambda \frac{1}{\meas(\Lambda)}
\sum_{\substack{\mathbf{j}\in\Z^\nu \\ \mathbf{j}\in\Lambda}}
\E\{\cF_{\omega,\Delta_{\mathbf{j}}}^+(t)\}\\
&\leq& \sup_\Lambda \sup_{\substack{\mathbf{j}\in\Z^\nu \\
\mathbf{j}\in\Lambda}}\E\{\cF_{\omega,\Delta_{\mathbf{j}}}^+(t)\}
= \sup_{\mathbf{j}\in \Z^\nu}\E\{\cF_{\omega,\Delta_{\mathbf{j}}}^+(t)\}.
\end{eqnarray*}
By metrical transitivity
\begin{displaymath}
\E\{\cF_{\omega,\Delta_{\mathbf{j}}}(t) \}= \E\{\cF_{T_{-\mathbf{j}}\omega,\Delta_0}(t) \}=
\E\{\cF_{\omega,\Delta_0}(t) \}.
\end{displaymath}
Further we estimate
\begin{eqnarray*}
|\cF_{\omega,\Delta_0}(t)| &=&
\left|\tr\left(e^{-t(H_0+\alpha_0(\omega)f)}-e^{-tH_0}
\right)\right|
\leq \left\|e^{-t(H_0+\alpha_0(\omega)f)}-e^{-tH_0} \right\|_{\cJ_1}.
\end{eqnarray*}
By Theorem \ref{thm:Stollmann:1} and Remark \ref{rem:Stollmann:1} this norm can
be bounded by
\begin{displaymath}
2^{2-\nu/4}(\pi t)^{-\nu/2}\left\|e^{-t(H_0+2W)/2}
\right\|^{1/2}_{\infty,\infty}\
\left\|e^{-t(H_0+4W)/4}\right\|^{1/2}_{\infty,\infty}
\end{displaymath}
with $W(x)=\min\{0,\alpha_- f_+(x),\alpha_+ f_-(x)\}$. Therefore for every
$t>0$ the quantities \newline
$\sup_{\mathbf{j}\in\Z^\nu}\E\{\cF_{\omega,\Delta_{\mathbf{j}}}^+(t)\}$ are
bounded and $\Gamma_-<\infty$. Similarly we can prove that $\Gamma_+>-\infty$.

Thus by the Akcoglu--Krengel ergodic theorem we obtain that for every $t>0$ the
limits
\begin{displaymath}
\lim_{\Lambda\rightarrow\infty}(\meas(\Lambda))^{-1}\cF_{\omega,\Delta_{\mathbf{j}}}^+(t)
\quad \mathrm{and}\quad
\lim_{\Lambda\rightarrow\infty}(\meas(\Lambda))^{-1}\cF_{\omega,\Delta_{\mathbf{j}}}^-(t)
\end{displaymath}
exist almost sure and are non-random. Thus we proved the first part of the
following

\begin{theorem}\label{SSDens:1}
For any $t>0$ the limit
\begin{displaymath}
\lim_{\Lambda\rightarrow\infty}\frac{1}{\meas(\Lambda)}
\int_\R e^{-t\lambda}\xi(\lambda; H_0+V_{\omega,\Lambda},H_0)d\lambda
\end{displaymath}
exists almost surely and is non-random. Moreover the integrated density of
states $N(\lambda)$ exists and the above limits equals
\begin{displaymath}
\int_\R e^{-t\lambda}(N_0(\lambda)-N(\lambda))d\lambda.
\end{displaymath}
\end{theorem}

The second part of the theorem follows from the estimates of Corollary
\ref{thm:Stollmann:3.2}.

If $f$ is sign-definite (say $f\geq 0$) and either all $\alpha_{\mathbf{j}}\geq
0$ or $\alpha_{\mathbf{j}}\leq 0$ there is a simpler proof of Theorem
\ref{SSDens:1}. From the inequality
\begin{displaymath}
1-e^{-(a+b)}\leq (1-e^{-a})+(1-e^{-b}),\quad ab\geq 0
\end{displaymath}
by the Feynman-Kac formula (see \cite{Geisler:Kostrykin:Schrader} for details)
it follows that
\begin{displaymath}
\cF_{\omega,\Lambda}(t)\leq \cF_{\omega,\Lambda_1}(t)+\cF_{\omega,\Lambda_2}(t)
\end{displaymath}
for all $t>0$. By the monotonicity property of the spectral shift function with
respect to the perturbation \cite{Birman:Yafaev,Geisler:Kostrykin:Schrader}
$\cF_{\omega,\Lambda}(t)\geq 0$ if $\alpha_{\mathbf{j}}(\omega)\leq 0$. If
$\alpha_{\mathbf{j}}(\omega)\geq 0$ then by Theorem \ref{thm:Stollmann:1} and
Lemma \ref{thm:Stollmann:2} we have
\begin{displaymath}
\inf_\Lambda (\meas(\Lambda))^{-1}\E\{\cF_{\omega,\Lambda}(t)\}>-\infty.
\end{displaymath}
Thus $\cF_{\omega,\Lambda}(t)$ satisfies the conditions of the Akcoglu--Krengel
theorem.

\begin{corollary}\label{SSDens:2}
For all $g\in C_0^1$ the limit
\begin{displaymath}
\lim_{\Lambda\rightarrow\infty}\frac{1}{\meas(\Lambda)}
\int g(\lambda)\xi(\lambda; H_0+V_{\omega,\Lambda},H_0)d\lambda =: \mu_\xi(g)
\end{displaymath}
exists almost surely and is non-random. Moreover
\begin{displaymath}
\mu_\xi(g)=\int g(\lambda)(N_0(\lambda)-N(\lambda))d\lambda.
\end{displaymath}
\end{corollary}

More precisely Corollary \ref{SSDens:2} states that there is a set
$\Omega_1\subseteq\Omega$ of full measure such that for all $\omega\in\Omega_1$
the limits exist for any $g$.

\begin{proof}
As in the proof of Theorem \ref{th.SSD.1} given $g\in C_0^1$ we approximate
$g(\lambda)$ by polynomials $P_k(\lambda)$ in $e^{-\lambda}$ such that
\begin{displaymath}
\sup_{\lambda\in A}e^{\lambda}|g(\lambda)-P_k(\lambda)|\rightarrow 0,\quad
A=\bigcup_\Lambda \spec(H_0+V_\Lambda)
\end{displaymath}
as $k\rightarrow\infty$. Then
\begin{eqnarray*}
\left|\int g(\lambda)\xi(\lambda; H_0+V_{\omega,\Lambda}, H_0)d\lambda
-\int P_k(\lambda)\xi(\lambda; H_0+V_{\omega,\Lambda},H_0)d\lambda \right|\\
\leq \int e^\lambda |g(\lambda)-P_k(\lambda)|\cdot e^{-\lambda}
|\xi(\lambda; H_0+V_{\omega,\Lambda},H_0)|d\lambda\\
\leq \|F_k\|_{L^\infty}\left\| e^{-(H_0+V_{\omega,\Lambda})}-e^{-H_0}\right\|_{\cJ_1},
\end{eqnarray*}
where $F_k=e^{\lambda}(g(\lambda)-P_k(\lambda))$. By Theorem
\ref{thm:Stollmann:1} and Lemma \ref{thm:Stollmann:2} it follows that
\begin{displaymath}
\left|\int g(\lambda)\frac{\xi(\lambda; H_0+V_{\omega,\Lambda}, H_0)}{\meas(\Lambda)}d\lambda
-\int P_k(\lambda)\frac{\xi(\lambda; H_0+V_{\omega,\Lambda},H_0)}{\meas(\Lambda)}d\lambda
\right|\leq C\|F_k\|_{L^\infty}
\end{displaymath}
with some $C>0$ independent of $\Lambda$ and $k$. By Theorem \ref{SSDens:1}
there is $\Omega_1\subseteq\Omega$ of full measure such that for any
$\omega\in\Omega_1$ the limit
\begin{displaymath}
\lim_{\Lambda\rightarrow\infty}(\meas(\Lambda))^{-1}
\int P_k(\lambda)\xi(\lambda; H_0+V_{\omega,\Lambda},H_0)d\lambda
\end{displaymath}
exists and is non-random for any finite $k\in\N$. Therefore
\begin{displaymath}
\left|\ulim_{\Lambda\rightarrow\infty}\int g(\lambda)
\frac{\xi(\lambda; H_0+V_{\omega,\Lambda}, H_0)}{\meas(\Lambda)}d\lambda-
\llim_{\Lambda\rightarrow\infty}\int g(\lambda)
\frac{\xi(\lambda; H_0+V_{\omega,\Lambda}, H_0)}{\meas(\Lambda)}d\lambda \right|
\leq 2\|F_k\|_{L^\infty},
\end{displaymath}
which proves the first part of the claim. The arguments used above in the proof
of Theorem \ref{th.SSD.1} give that if $g\in C_0^2$ then the relation
\begin{displaymath}
\lim_{\Lambda\rightarrow\infty}\frac{1}{\meas(\Lambda)}
\int g(\lambda)\xi(\lambda; H_0+V_{\omega,\Lambda},H_0)d\lambda
=\int g(\lambda)(N_0(\lambda)-N(\lambda))d\lambda
\end{displaymath}
holds almost surely.
\end{proof}

Recall that if $\lambda<0$ then
$\xi(\lambda;H_0+V_{\omega,\Lambda},H_0)=-N(\lambda;H_0+V_{\omega,\Lambda})$,
the eigenvalue counting function for the operator $H_0+V_{\omega,\Lambda}$.

\begin{corollary}\label{SSDens:3}
The relation
\begin{displaymath}
\lim_{\Lambda\rightarrow\infty}\frac{1}{\meas(\Lambda)}
\xi(\lambda; H_0+V_{\omega,\Lambda},H_0)=-N(\lambda)
\end{displaymath}
is valid almost surely for all $\lambda<0$ which are continuity points of
$N(\lambda)$.
\end{corollary}

\begin{proof}
The proof is standard (see e.g.\ \cite{Pastur:71,Nakao}). Since the
one-dimensional case was treated in detail in \cite{Kostrykin:Schrader:98a} we
consider the case $\nu\geq 2$ only. From Corollary \ref{SSDens:2} it follows
that for any $g\in C_0^2$ supported in $(-\infty,0)$
\begin{equation}\label{SSDens:3:eq1}
\lim_{\Lambda\rightarrow\infty}\int g(\lambda) d\xi_{\omega,\Lambda}(\lambda)=
-\int g(\lambda) dN(\lambda)
\end{equation}
almost surely, where
\begin{displaymath}
\xi_{\omega,\Lambda}(\lambda)=(\meas(\Lambda))^{-1}\xi(\lambda;H_0+V_{\omega,\Lambda},H_0).
\end{displaymath}
For $\lambda<0$ by the Cwieckel-Lieb-Rosenblum estimate (see e.g. \cite{RS4})
for $\nu\geq 3$
\begin{eqnarray}\label{SSDens:3:eq2}
\lefteqn{-\xi_{\omega,\Lambda}(\lambda)\leq C (\meas(\Lambda))^{-1}\int_{\R^\nu}
|(V_{\omega,\Lambda}(x))_-|^{\nu/2}dx} \nonumber\\ &\leq&
C|\min\{0,\alpha_-\}|^{\nu/2}
\|f_+\|_{L^{\nu/2}}^{\nu/2} + C(\max\{0,\alpha_+\})^{\nu/2}
\|f_-\|_{L^{\nu/2}}^{\nu/2}
\end{eqnarray}
with some uniform constant $C>0$. For $\nu=2$ by Proposition 6.1 of
\cite{Birman:Solomyak:92}
\begin{eqnarray}\label{SSDens:3:eq3}
\lefteqn{-\xi_{\omega,\Lambda}(\lambda)\leq C (\meas(\Lambda))^{-1}
\|(V_{\omega,\Lambda})_-\|_{l^1(L^\sigma)}}
\nonumber\\
&\leq& C |\min\{0,\alpha_-\}|
\|f_+\|_{L^{\sigma}} + C \max\{0,\alpha_+\}
\|f_-\|_{L^{\sigma}}
\end{eqnarray}
for any $\sigma>1$. Note that the quantities on the r.h.s. of
\eqref{SSDens:3:eq2} and
\eqref{SSDens:3:eq3} are finite. Indeed for $\nu\geq 4$ any compactly supported
function $f\in L^p(\R^\nu)$ with some $p>\nu/2$ belongs also to
$L^{\nu/2}(\R^\nu)$. Similarly in the case $\nu\leq 3$ any square integrable
$f$ with compact support belongs to $L^p(\R^\nu)$ with arbitrary $1\leq p\leq
2$.

Since $\xi_{\omega,\Lambda}(\lambda)$ are monotone functions these estimates
imply that for every $\omega\in\Omega$ the family
$\{\xi_{\omega,\Lambda}(\lambda)\}_\Lambda$ is of uniformly bounded variation
on $(-\infty,0)$. By Helly's Selection Theorem for every $\omega\in\Omega$
there is a sequence $\Lambda_i$, $i=1,2,\ldots$ such that
$\lim_{i\rightarrow\infty}\xi_{\omega,\Lambda_i}(\lambda)=\xi^{(\omega)}(\lambda)$
for all those $\lambda\in(-\infty,0)$ which are continuity point of
$\xi^{(\omega)}(\lambda)$. By Helly's second theorem it follows from this that
\begin{displaymath}
\lim_{i\rightarrow\infty}\int g(\lambda) d\xi_{\omega,\Lambda_i}(\lambda)=
\int g(\lambda) d\xi^{(\omega)}(\lambda)
\end{displaymath}
for any $\omega\in\Omega$ and any $g\in C_0^2$ with support in $(-\infty,0)$.
From \eqref{SSDens:3:eq1} it follows that
\begin{displaymath}
\int g(\lambda) d\xi^{(\omega)}(\lambda) = -\int g(\lambda) dN(\lambda)
\end{displaymath}
for $\P$-almost all $\omega\in\Omega$ and all $g\in C_0^2$. Hence
$\xi^{(\omega)}(\lambda)=-N(\lambda)+C$ a.e.\ with some constant $C$ for
$\P$-almost all $\omega\in\Omega$. But $\xi^{(\omega)}(\lambda)=-N(\lambda)=0$
for sufficiently large negative $\lambda$ and thus $C=0$. Now we note that two
monotone functions which are equal almost everywhere can be different only at
the points of discontinuity. This remark completes the proof of the corollary.
\end{proof}

\subsection{Random Potential Concentrated near a Hyperplane}

Consider a decomposition $\Z^\nu=\Z^{\nu_1}\oplus\Z^{\nu_2}$ with
$\nu_1+\nu_2=\nu$, $\nu_1,\nu_2\geq 1$. Let
\begin{equation}\label{ran:pot:surf}
V_\omega(x) = \sum_{\mathbf{j}\in \Z^{\nu_1}} \alpha_{\mathbf{j}}(\omega)
f(x-\mathbf{j}).
\end{equation}
Let now $\Lambda_1$ be a box in $\R^{\nu_1}\subset\R^\nu$ and we approximate
$V_\omega$ by
\begin{equation}\label{ran:pot:surf1}
V_{\omega,\Lambda_1}(x) = \sum_{\substack{\mathbf{j}\in \Z^{\nu_1} \\
\mathbf{j}\in\Lambda_1}}
\alpha_{\mathbf{j}}(\omega) f(x-\mathbf{j}).
\end{equation}

As for the case of the lattice $\Z^\nu$ we have

\begin{proposition}\label{SSDens:4}
For any $t>0$ the limit
\begin{displaymath}
\lim_{\Lambda_1\rightarrow\infty}\frac{1}{\meas_{\nu_1}(\Lambda_1)}
\int_\R e^{-t\lambda}\xi(\lambda; H_0+V_{\omega,\Lambda_1},H_0) d\lambda=: \cL(t)
\end{displaymath}
exists almost surely and is non-random.
\end{proposition}

The proof is completely analogous to that of Theorem \ref{SSDens:1} and
therefore will be omitted.

\begin{corollary}\label{SSDens:5}
For all $g\in C_0^1$ the limit
\begin{equation}\label{SSDens:5:equ}
\lim_{\Lambda\rightarrow\infty}\frac{1}{\meas_{\nu_1}(\Lambda)}
\int_\R g(\lambda)\xi(\lambda; H_0+V_{\omega,\Lambda}, H_0)d\lambda=:\mu(g)
\end{equation}
exists almost surely and is non-random. The linear functional $\mu(g)$ defines
a distribution (of order 1) $\xi(\lambda)$ such that
\begin{displaymath}
\mu(g)=\int_\R g(\lambda)\xi(\lambda) d\lambda.
\end{displaymath}
Moreover $\mu(g)$ is related to the density of surface states functional
$\mu_s(g)$ (see \cite{Englisch:Kirsch:Schroeder:Simon,Chahrour}) such that
$\mu_s(g)=\mu(g')$, where
\begin{displaymath}
\mu_s(g)= \lim_{\substack{\Lambda_1\rightarrow\infty\\\Lambda_2\rightarrow\infty}}
\frac{1}{\meas_{\nu_1}(\Lambda_1)}\tr\left[\chi_{\Lambda_1\times\Lambda_2}
(g(H_0+V_{\omega,\Lambda_1})-g(H_0))\right],\quad g\in C_0^2,
\end{displaymath}
almost surely for arbitrary sequences of boxes $\Lambda_1\subset\R^{\nu_1}$,
$\Lambda_2\subset\R^{\nu_2}$ tending to infinity.
\end{corollary}

\begin{remark}
More precisely Corollary \ref{SSDens:5} asserts that there is a set
$\Omega_1\subseteq\Omega$ of full measure such that for all $\omega\in\Omega_1$
the limits exist for any $g$.
\end{remark}

The almost surely existence of the limit \eqref{SSDens:5:equ} follows from
Proposition \ref{SSDens:4}. To prove the second part of the claim it suffices
to show that
\begin{displaymath}
-t\cL(t)=\lim_{\substack{\Lambda_1\rightarrow\infty \\ \Lambda_2\rightarrow\infty}}
\frac{1}{\meas_{\nu_1}(\Lambda_1)}\tr\left[\chi_{\Lambda_1\times\Lambda_2}
 \left(e^{-t(H_0+V_{\omega,\Lambda_1})}-e^{-tH_0} \right)\right].
\end{displaymath}
In turn this follows immediately from the following

\begin{lemma}\label{SSDens:5:1}
Let $\Lambda=\Lambda_1\times\Lambda_2$ be a box such that
$\Lambda_1\subset\R^{\nu_1}$, $\Lambda_2\subset\R^{\nu_2}$.  If $\nu_1\geq 2$
then for every $t>0$ there are constants $c_1,c_2>0$ such that
\begin{eqnarray*}
\left\|\chi_\Lambda\left(e^{-t(H_0+V_\omega)}-
e^{-t(H_0+V_{\omega,\Lambda_1})}\right)\right\|_{\cJ_1}
\leq c_1\ \meas_{\nu_1-1}(\partial\Lambda_1),\\
\left\|(1-\chi_\Lambda)\left(e^{-t(H_0+V_{\omega,\Lambda_1})}-
e^{-tH_0}\right)\right\|_{\cJ_1}\leq c_2\ \meas_{\nu_1-1}(\partial\Lambda_1)
\end{eqnarray*}
for all $\omega\in\Omega$. If $\nu_1=1$ the same inequalities hold if their
r.h.s.\ are replaced by some constants.
\end{lemma}

\begin{proof}
Let $\Lambda_1^c$ denote the complement of $\Lambda_1$ in $\R^{\nu_1}$. Also we
denote $\Lambda'_1=\Lambda_1\times[-1/2,1/2]^{\nu_2}$ and
$(\Lambda_1^c)^\prime=\Lambda_1^c\times[-1/2,1/2]^{\nu_2}$. Now we write
\begin{eqnarray*}
\lefteqn{\chi_\Lambda\left(e^{-t(H_0+V_\omega)}-e^{-t(H_0+V_{\omega,\Lambda_1})}\right)}\\
&=&
\chi_\Lambda\left(e^{-t(H_0+V_\omega)}-e^{-t(H_0+V_\omega+\infty_{(\Lambda_1^c)^\prime})}\right)
-\chi_\Lambda\left(e^{-t(H_0+V_{\omega,\Lambda_1})}-
e^{-t(H_0+V_{\omega,\Lambda_1}+\infty_{(\Lambda_1^c)^\prime})}\right).
\end{eqnarray*}
Repeating the arguments used in the proof of Lemma  \ref{thm:Stollmann:5} we
obtain that both
\begin{displaymath}
\left\|\chi_\Lambda\left(e^{-t(H_0+V_\omega)}-
e^{-t(H_0+V_\omega+\infty_{(\Lambda_1^c)^\prime})}\right)\right\|_{\cJ_1}
\end{displaymath}
and
\begin{displaymath}
\left\|\chi_\Lambda\left(e^{-t(H_0+V_{\omega,\Lambda_1})}-
e^{-t(H_0+V_{\omega,\Lambda_1}+\infty_{(\Lambda_1^c)^\prime}}
\right)\right\|_{\cJ_1}
\end{displaymath}
are bounded by
\begin{eqnarray*}
\lefteqn{2^{1-\nu/4}(\pi
t)^{-\nu/2}\left\|e^{-t(H_0+2(V_\omega)_-)/2}\right\|_{\infty,\infty}^{1/2}\
\left\|e^{-t(H_0+4(V_\omega)_-)/4}\right\|_{\infty,\infty}^{1/2}}\\
&&\cdot\left(\|\chi_\Lambda\mathbf{P}_\bullet\{\tau_{(\Lambda_1^c)^\prime}\leq
t/2\}^{1/2}\|_{L^1}
+\|\mathbf{E}_\bullet\{\chi_\Lambda(X_t);\tau_{(\Lambda_1^c)^\prime}\leq
t/2\}^{1/2}\|_{L^1} \right)\\ &\leq& 2^{1-\nu/4}(\pi
t)^{-\nu/2}\left\|e^{-t(H_0+2V^{(-)})/2}\right\|_{\infty,\infty}^{1/2}\
\left\|e^{-t(H_0+4V^{(-)})/4}\right\|_{\infty,\infty}^{1/2}\\
&&\cdot\left(\|\chi_\Lambda\mathbf{P}_\bullet\{\tau_{(\Lambda_1^c)^\prime}\leq
t/2\}^{1/2}\|_{L^1}
+\|\mathbf{E}_\bullet\{\chi_\Lambda(X_t);\tau_{(\Lambda_1^c)^\prime}\leq
t/2\}^{1/2}\|_{L^1} \right),
\end{eqnarray*}
where $V^{(-)}=\min\{0,\alpha_-\}\sum_{\mathbf{j}\in\Z^{\nu_1}}
f_+(\cdot-\mathbf{j})+\max\{0,\alpha_+\}\sum_{\mathbf{j}\in\Z^{\nu_1}}
f_-(\cdot-\mathbf{j})$. By Lemmas \ref{thm:Stollmann:2} and
\ref{thm:Stollmann:3.1} the expression in the brackets can be bounded by a
constant times $\meas_{\nu_1-1}(\partial\Lambda_1)$ if $\nu_1\geq 2$ and simply
by a constant if $\nu_1=1$. The second inequality in the claim of the lemma can
be proved similarly.
\end{proof}

\begin{corollary}\label{SSDens:5:2}
For $\lambda<0$ the limit
\begin{displaymath}
\lim_{\Lambda\rightarrow\infty}\frac{1}{\meas_{\nu_1}(\Lambda)}
\xi(\lambda; H_0+V_{\omega,\Lambda},H_0)=: -N(\lambda)
\end{displaymath}
exists almost surely in all points of continuity of the non-decreasing function
$N(\lambda)$ and is non-random.
\end{corollary}

\begin{remark}\label{rem:SSDens:5:2}
By Corollary \ref{SSDens:5} $N(\lambda)$ is the integrated density of surface
states.
\end{remark}

\textit{A priori} in the general case it is not clear whether the sign-indefinite
functional $\mu(g)$ defines some signed measure rather than a distribution. If
we could prove that $\mu(g)$ is continuous on continuous functions of compact
support then we would be able to show that $\mu(g)=\mu_+(g)-\mu_-(g)$ with
$\mu_\pm(g)$ being some positive linear functionals (see e.g.\ Theorem IV.16 in
\cite{RS1}), and thus by Riesz's representation theorem will define a signed
Borel measure. We will not discuss the continuity of $\mu(g)$ in the general
case. Instead we will suppose that the single-site potential is non-negative,
$f\geq 0$.

\begin{lemma}\label{SSDens:6}
Let $\{\alpha_{\mathbf{j}}(\omega)\}_{\mathbf{j}\in \Z^{\nu_1}}$ be a sequence
of i.i.d.\ variables forming a stationary, metrically transitive random field.
Then $\alpha^+_{\mathbf{j}}(\omega)=\max\{\alpha_{\mathbf{j}}(\omega),0\}$ and
$\alpha^-_{\mathbf{j}}(\omega)=\min\{\alpha_{\mathbf{j}}(\omega),0\}$ are
sequences of i.i.d.\ variables which also form stationary, metrically
transitive fields.
\end{lemma}

Indeed
$\alpha^+_{\mathbf{j}}(T_{\mathbf{k}}\omega)=
\max\{\alpha_{\mathbf{j}}(T_{\mathbf{k}}\omega),0\} =
\max\{\alpha_{\mathbf{j}-\mathbf{k}}(\omega),0\}=\alpha^+_{\mathbf{j}-\mathbf{k}}(\omega)$
and similarly
$\alpha^-_{\mathbf{j}}(T_{\mathbf{k}}\omega)=
\alpha^-_{\mathbf{j}-\mathbf{k}}(\omega)$.

\begin{remark}
The distributions $\kappa^\pm$ of
$\{\alpha^{\pm}_{\mathbf{j}}(\omega)\}_{\mathbf{j}\in \Z^{\nu_1}}$ can be
expressed in terms of the distribution $\kappa$ of
$\{\alpha_{\mathbf{j}}(\omega)\}_{\mathbf{j}\in \Z^{\nu_1}}$. If $\kappa$ is
concentrated on a subset of $[0,\infty)$ then $\kappa^+=\kappa$ and
$\kappa^-=0$. Otherwise $\kappa^+=\kappa|_{\R_+}+\kappa_0$, where
$\kappa|_{\R_+}$ is the restriction of the measure $\kappa$ to the non-negative
semiaxis and $\kappa_0$ is a point measure concentrated at zero such that
$\kappa_0(\{0\})=\kappa((-\infty,0))$. The measure $\kappa^-$ can be described
similarly.
\end{remark}

\begin{proposition}\label{SSDens:7}
Suppose that $f\geq 0$ and either all $\alpha_{\mathbf{j}}\geq 0$ or all
$\alpha_{\mathbf{j}}\leq 0$. Then the linear functional $\mu(g)$
(\ref{SSDens:5:equ}) induces a positive (negative) Borel measure
$d\Xi(\lambda)$ such that
\begin{displaymath}
\mu(g)=\int_\R g(\lambda) d\Xi(\lambda).
\end{displaymath}
Moreover for all $\lambda\in\R$ the limit
\begin{displaymath}
\lim_{\Lambda\rightarrow\infty}(\meas_{\nu_1}(\Lambda))^{-1}\int_{-\infty}^\lambda\xi(E; H_0+V_{\omega,\Lambda},H_0)
dE
\end{displaymath}
exists almost surely and equals $\Xi(\lambda)=\Xi((-\infty,\lambda))$ for every
continuity point of $\Xi(\lambda)$.
\end{proposition}

\begin{proof}
We consider the case $\alpha_{\mathbf{j}}\geq 0$ only since the proof for the
case $\alpha_{\mathbf{j}}\leq 0$ carries over verbatim. By the monotonicity
property of the spectral shift function
\cite{Birman:Yafaev,Geisler:Kostrykin:Schrader} $\xi(\lambda;
H_0+V_{\omega,\Lambda}, H_0)\geq 0$ for Lebesgue almost all $\lambda\in\R$, all
$\omega\in\Omega$ and all $\Lambda$. From this it follows that the functional
$\mu(g)$ is positive. As it is noted in \cite{Englisch:Kirsch:Schroeder:Simon}
Riesz's representation theorem extends to the case of linear positive
functionals on $C_0^k$ and thus defines a positive Borel measure
$d\Xi(\lambda)$.
\end{proof}

Finally we consider the case with no restriction on the sign of the
$\alpha_{\mathbf{j}}$'s.

\begin{theorem}\label{SSDens:8}
Let $f\geq 0$. Then the linear functional $\mu(g)$ (\ref{SSDens:5:equ}) induces
a signed Borel measure $d\Xi(\lambda)$ such that
\begin{displaymath}
\mu(g)=\int_\R g(\lambda)d\Xi(\lambda).
\end{displaymath}
Moreover for all $\lambda\in\R$ the limit
\begin{displaymath}
\lim_{\Lambda\rightarrow\infty}(\meas_{\nu_1}(\Lambda))^{-1}\int_{-\infty}^\lambda\xi(E; H_0+V_{\omega,\Lambda},H_0)
dE
\end{displaymath}
exists almost surely and equals á function of locally bounded variation
$\Xi(\lambda)=\Xi((-\infty,\lambda))$ for every continuity point of
$\Xi(\lambda)$.
\end{theorem}

\begin{proof}
For almost every $\lambda\in\R$, every $\omega\in\Omega$ and for arbitrary
$\Lambda$ by the chain rule for the spectral shift function (see e.g.\
\cite{Birman:Yafaev}) we have
\begin{equation}\label{SSDens:8:eq}
\xi(\lambda;H_0+V_{\omega,\Lambda}, H_0) = \xi(\lambda; H_0+V_{\omega,\Lambda}^+
+V_{\omega,\Lambda}^-,
H_0+V_{\omega,\Lambda}^-)+\xi(\lambda;H_0+V_{\omega,\Lambda}^-,H_0)
\end{equation}
with
$V_{\omega,\Lambda}^\pm=\sum_{\mathbf{j}\in\Lambda}\alpha^\pm_{\mathbf{j}}(\omega)
 f(\cdot-\mathbf{j})$. Here $\alpha_{\mathbf{j}}^\pm(\omega)$ is the decomposition of
$\alpha_{\mathbf{j}}(\omega)$ into its positive and negative part such that
$V_{\omega,\Lambda}=V_{\omega,\Lambda}^++V_{\omega,\Lambda}^-$. By the
monotonicity property of the spectral shift function we have that the first
summand on the r.h.s. of (\ref{SSDens:8:eq}) is a.e.\ non-negative and the
second one is a.e.\ non-positive. By Corollary \ref{SSDens:5} there is a linear
functional
\begin{displaymath}
\mu(g)=\lim_{\Lambda\rightarrow\infty}(\meas_{\nu_1}(\Lambda))^{-1}
\int_\R g(\lambda) \xi(\lambda; H_0+V_{\omega,\Lambda}, H_0)d\lambda.
\end{displaymath}
By Lemma \ref{SSDens:6} there is a negative linear functional which we denote
by $\mu^-(g)$ such that
\begin{displaymath}
\mu^-(g)=\lim_{\Lambda\rightarrow\infty}(\meas_{\nu_1}(\Lambda))^{-1}
\int_\R g(\lambda) \xi(\lambda; H_0+V_{\omega,\Lambda}^-, H_0)d\lambda.
\end{displaymath}
By (\ref{SSDens:8:eq}) the limit
\begin{displaymath}
\lim_{\Lambda\rightarrow\infty}(\meas_{\nu_1}(\Lambda))^{-1}
\int_\R g(\lambda) \xi(\lambda; H_0+V_{\omega,\Lambda}^-+V_{\omega,\Lambda}^+,
H_0+V_{\omega,\Lambda}^-)d\lambda.
\end{displaymath}
exists almost surely and defines a non-random linear positive functional which
we denote by $\mu^+(g)$. Thus
\begin{displaymath}
\mu(g)=\mu^+(g)+\mu^-(g),
\end{displaymath}
i.e. is a difference of two positive linear functionals and therefore defines a
signed Borel measure $d\Xi(\lambda)$.
\end{proof}

The existence of the spectral shift function in the sense of distribution for
the discrete Schr\"{o}\-din\-ger operators (Jacobi matrices) with potentials of the
type (\ref{ran:pot:surf}) was proved by A. Chahrour in \cite{Chahrour}. Theorem
\ref{SSDens:8} improves this result, i.e.\ we prove that the spectral shift
density is defined as a measure rather than a distribution of order 1.

%\newpage

%%%%%%%%%%%%%%%%%%%%%%%%%%%%%%%%%%%%%%%%%%%%%%%%%%%%%%
\section*{Appendix}
\renewcommand{\theequation}{A.\arabic{equation}}
\renewcommand{\thesection}{A}
\setcounter{equation}{0}
\setcounter{theorem}{0}
\setcounter{subsection}{0}

In this appendix for the convenience of the reader we collect some well known
technical facts used in this article.

%%%%%%%%%%%%%%%%%%%%%%%%%%%%%%%%%%%%%%%%%%%%%%%%%%%%%%
\subsection{Schr\"{o}dinger Semigroup Estimates}
%%%%%%%%%%%%%%%%%%%%%%%%%%%%%%%%%%%%%%%%%%%%%%%%%%%%%%

The Feynman-Kac formula gives

\begin{theorem}\label{SSG1}
Let $V_1$, $V_2$ be such that ${V_1}_+,{V_2}_+\in K_\nu^{\mathrm{loc}}$,
${V_1}_-,{V_2}_-\in K_\nu$ and $V_1\geq V_2$. Then for all $f\geq 0$ and $0\leq
f\in L^p(\R^\nu)$ with $p\geq 1$
\begin{displaymath}
0\leq e^{-t(H_0+V_1)}f\leq e^{-t(H_0+V_2)}f
\end{displaymath}
almost everywhere.
\end{theorem}

The next result is a special case of hypercontractivity properties of
Schr\"{o}dinger semigroups:

\begin{theorem}\label{A.2:neu}\cite{Simon:82}
Let $V_-\in K_\nu$, $V_+\in K_\nu^{loc}$. Then for every $t>0$ and $p\leq q$
the operator $e^{-tH}$ is bounded from $L^p$ to $L^q$ and
\begin{displaymath}
\left\|e^{-tH}\right\|_{2,2}\leq \left\|e^{-tH}\right\|_{p,p}\leq
\left\|e^{-tH}\right\|_{\infty,\infty}
\end{displaymath}
for any $p\geq 2$.
\end{theorem}

Since $\|e^{-t(H_0+V)}\|_{\infty,\infty}=\|e^{-t(H_0+V)}1\|_{L^{\infty}}$
Theorem \ref{SSG1} implies the following monotonicity property of the norm with
respect to the potential $V$
\begin{equation}\label{SSG2}
\left\|e^{-t(H_0+V_1)}\right\|_{\infty,\infty}\leq
\left\|e^{-t(H_0+V_2)}\right\|_{\infty,\infty}.
\end{equation}
For all $1\leq p\leq q \leq \infty$ and arbitrary $A>-\inf\spec(H)\geq 0$ there
is a constant $C_{p,q}$ such that the inequality
\begin{equation}\label{SSG3}
\left\|e^{-tH}\right\|_{p,q}\leq C_{p,q} t^{-\gamma} e^{At}
\end{equation}
holds with $\gamma=\nu(p^{-1}-q^{-1})/2$. The proof of (\ref{SSG3}) is given in
\cite{Simon:82}.
From Theorem \ref{A.2:neu} it follows (see \cite{Simon:82} for details) that
$e^{-tH}$ is an integral operator and
\begin{displaymath}
\left\|e^{-tH} \right\|_{p,\infty}=
\sup_x\left\{\int\left(e^{-tH}(x,y)\right)^q dy \right\}^{1/q},
\end{displaymath}
where $q^{-1}=1-p^{-1}$ for any $1\leq p <\infty$.

\begin{lemma}\label{A.3:neu}\cite{Simon:82}
Let $V$ be such that $V_-\in K_\nu$ and $V_+\in K_\nu^{\mathrm{loc}}$. Then for
all $t>0$
\begin{displaymath}
\left\|e^{-t(H_0+V)} \right\|_{2,\infty}^2 \leq (4\pi t)^{-\nu/2}\left\|e^{-t(H_0+2V)}\right\|_{\infty,\infty}.
\end{displaymath}
\end{lemma}

\begin{proof}
Using the Schwarz inequality with respect to the Wiener measure in the
Feynman-Kac formula we obtain
\begin{equation}\label{SSG4}
|(e^{-t(H_0+V)}f)(x)|\leq \left[\left(e^{-t(H_0+2V)}1\right)(x)\right]^{1/2}\
\left[\left(e^{-tH_0}|f|^2\right)(x)\right]^{1/2}
\end{equation}
for any $f\in L^2$. The operator $e^{-tH_0}$ is convolution by the function
$(4\pi t)^{-\nu/2}\exp(-x^2/4t)$. Since this function is in $L^\infty$, by the
Young inequality we have
\begin{displaymath}
\left\|e^{-tH_0}g \right\|_{L^\infty}\leq (4\pi t)^{-\nu/2} \|g\|_{L^1}
\end{displaymath}
with $g=|f|^2$. Therefore by (\ref{SSG4})
\begin{displaymath}
\left\|e^{-t(H_0+V)}f\right\|^2_{L^\infty}\leq (4\pi t)^{-\nu/2}
\left\|e^{-t(H_0+2V)} \right\|_{\infty,\infty} \|f\|_{L^2}^2,
\end{displaymath}
thus proving the lemma.
\end{proof}

%%%%%%%%%%%%%%%%%%%%%%%%%%%%%%%%%%%%%%%%%%%%%%%%%%%%%%
\subsection{Trace and Hilbert-Schmidt Norm Estimates}
%%%%%%%%%%%%%%%%%%%%%%%%%%%%%%%%%%%%%%%%%%%%%%%%%%%%%%

Here we collect some Hilbert-Schmidt and trace norm estimates. The following
lemmas are especially useful for estimating norms of semigroup differences and
are special cases of the ``little Grothendick theorem" \cite{Defant:Floret}.

\begin{lemma}\label{Stollmann:HS}\cite{Stollmann:94}
Let $A\in\cL(C(\R^\nu),L^2(\R^\nu))$, $B\in\cL(L^2(\R^\nu),C(\R^\nu))$ and
assume that $A$ preserves positivity (i.e. $f\geq 0$ implies $Af\geq 0$
pointwise). Then the operator $AB:\ L^2(\R^\nu)\rightarrow L^2(\R^\nu)$ is
Hilbert-Schmidt and
\begin{displaymath}
\|AB\|_{\cJ_2}\leq \|A\|_{\infty,2}\ \|B\|_{2,\infty}.
\end{displaymath}
\end{lemma}

\begin{lemma}\label{Stollmann:Tr}\cite{Stollmann:94,Demuth:St:St:Cas}
Let $A\in\cL(L^1(\R^\nu),L^2(\R^\nu)))$, $B\in\cL(L^2(\R^\nu),L^1(\R^\nu)))$
and let $B$ preserve positivity. Let also there is $\phi\in L^1(\R^\nu)$ such
that $|(Bf)(x)|\leq \phi(x)$ for all $f\in L^2$ with $\|f\|_{L^2}\leq 1$. Then
$AB\in\cJ_1$ and
\begin{displaymath}
\|AB\|_{\cJ_1}\leq \|A\|_{1,2}\ \|\phi\|_{L^1}.
\end{displaymath}
\end{lemma}


\newpage

\begin{thebibliography}{99}

\bibitem{Alama:Deift:Hempel} S. Alama, P.A. Deift, and R. Hempel,
\textit{Eigenvalue branches of the Schr\"{o}dinger operator $H-\lambda W$ in
a gap of $\sigma(H)$}, Commun. Math. Phys. \textbf{121}, 291 -- 321 (1989).

\bibitem{Barbaroux:Combes:Hislop:97} J.M. Barbaroux, J.M. Combes, and P.D.Hislop,
\textit{Localization near band edges for random Schr\"{o}dinger operators},
Helv. Phys. Acta, \textbf{70}, 16 -- 43 (1997).

\bibitem{Birman:Krein} M.Sh. Birman and M.G. Krein, \textit{On the theory of wave operators
and scattering operators}, Sov. Math.-Doklady \textbf{3}, 740 -- 744 (1962).

\bibitem{Birman:Solomyak:69} M.Sh. Birman and M.Z. Solomjak, \textit{Estimates of singular
numbers of integral operators. III. Operators in unbounded domains}, Vestnik
Leningrad. Univ. \textbf{24}, 35 -- 48 (1969) (in Russian).

\bibitem{Birman:Solomyak:92} M.Sh. Birman and M. Solomyak, \textit{Schr\"{o}dinger operator.
Estimates for number of bound states as function-theoretical problem}, Amer.
Math. Soc. Transl. (2) \textbf{150}, 1 -- 54 (1992).

\bibitem{Birman:Yafaev} M.Sh. Birman and D.R. Yafaev, \textit{The spectral shift
function. Work by M.G.Krein and its further development}, St. Petersburg Math.
J. \textbf{4}, 833 -- 870 (1993).

\bibitem{Birman:Pushnitski} M.Sh. Birman and A.B. Pushnitski, \textit{Spectral shift
function, amazing and multifaceted}, Integr. Eq. Oper. Theory \textbf{30}, 191
-- 199 (1998)

\bibitem{Chahrour} A. Chahrour, \textit{Sur la densit\'{e} int\'{e}gr\'{e}e d'\'{e}tats surfaciques
et la fonction g\'{e}n\'{e}ralis\'{e}e de d\'{e}placement spectral pour un potentiel surfacique
ergodique}, Helv. Phys. Acta (to appear).

\bibitem{Combes:Hislop:94} J.M. Combes and P.D. Hislop, \textit{Localization
for some continuous, random Hamiltonians in $d$-dimensions}, J. Funct. Anal.
\textbf{124}, 149 -- 180 (1994).

\bibitem{Cycon:Froese:Kirsch:Simon} H.L. Cycon, R.G. Froese, W. Kirsch, and B. Simon,
\textit{Schr\"{o}dinger Operators}, Springer, Berlin, 1987.

\bibitem{Defant:Floret} A. Defant and K. Floret, \textit{Tensor Norms and Operator Ideals},
North Holland, Amsterdam, 1993.

\bibitem{Demuth:91} M. Demuth, \textit{On large coupling operator norm convergences
of resolvent differences}, J. Math. Phys. \textbf{32}, 1522 -- 1530 (1991).

\bibitem{Demuth:St:St:Cas} M. Demuth, P. Stollmann, G. Stolz, and J. van Casteren,
\textit{Trace norm estimates for products of integral operators and diffusion
semigroups}, Integr. Eq. Oper. Theory, \textbf{23}, 145 -- 153 (1995).

\bibitem{Englisch:Kirsch:Schroeder:Simon} H. Englisch, W. Kirsch, M. Schr\"{o}der,
and B. Simon, \textit{Random Hamiltonians ergodic in all but one direction},
Commun. Math. Phys. \textbf{128}, 613 -- 625 (1990).

\bibitem{Geisler:Kostrykin:Schrader} R. Geisler, V. Kostrykin, and R. Schrader,
\textit{Concavity properties of Krein's
spectral shift function}, Rev. Math. Phys. \textbf{7}, 161 -- 181 (1995).

\bibitem{Gesztesy:Makarov:Naboko} F. Gesztesy, K.A. Makarov, and S.N. Naboko,
\textit{The spectral shift operator}, p.\ 59 -- 90 in J. Dittrich, P. Exner, and M. Tater (Eds.)
\textit{Mathematical Results in Quantum Mechanics}, Operator Theory: Advances and
Applications, Vol. 108, Birkh\"{a}user, Basel, 1999.

\bibitem{Gesztesy:Makarov} F. Gesztesy and K.A. Makarov, \textit{The $\Xi$ operator
and its relation to Krein's spectral shift operator}, preprint (1999);
available from \texttt{http://www.ma.utexas.edu/mp\_arc/}.

\bibitem{Fischer:Hupfer:Leschke:Mueller} W. Fischer, T. Hupfer, H. Leschke, and P. M\"{u}ller,
\textit{Existence of the density of states for multi-dimensional continuum Schr\"{o}dinger
operators with Gaussian random potentials}, Commun. Math. Phys. \textbf{190},
133 -- 141 (1997).

\bibitem{Hempel:92} R. Hempel, \textit{Eigenvalues in gaps and decoupling by
Neumann boundary conditions}, J. Math. Anal. Appl. \textbf{169}, 229 -- 259
(1992).

\bibitem{Hempel:Kirsch} R. Hempel and W. Kirsch, \textit{On the integrated density of
states for crystals with randomly distributed impurities}, Commun. Math. Phys.
\textbf{159}, 459 -- 469 (1994).

\bibitem{Keller} J.B. Keller, \textit{Discriminant, transmission coefficient, and stability
bands of Hill's equation}, J. Math. Phys. \textbf{25}, 2903 -- 2904 (1984).

\bibitem{Kirsch:Martinelli:82a} W. Kirsch and F. Martinelli, \textit{On the density of
states of Schr\"{o}dinger operators with a random potential}, J. Phys. A: Math.
Gen. \textbf{15}, 2139 -- 2156 (1982).

\bibitem{Kirsch:Kotani:Simon} W. Kirsch, S. Kotani, and B. Simon, \textit{Absence of
absolutely continuous spectrum for some one dimensional random but
deterministic operators}, Ann. Inst. Henri Poincar\'{e}, Phys. theor. \textbf{42},
383 -- 406 (1985).

\bibitem{Kirsch:87} W. Kirsch, \textit{Small perturbations and the eigenvalues
of the Laplacian on large bounded domains}, Proc. Amer. Math. Soc.
\textbf{101}, 509 -- 512 (1987).

\bibitem{Kostrykin:Schrader:94} V. Kostrykin and R. Schrader, \textit{Cluster properties of one
particle Schr\"{o}dinger operators}, Rev. Math. Phys. \textbf{6}, 833 -- 853
(1994).

\bibitem{Kostrykin:Schrader:98} V. Kostrykin and R. Schrader, \textit{Cluster properties of one
particle Schr\"{o}dinger operators. II}, Rev. Math. Phys. \textbf{10}, 627
-- 683 (1998).

\bibitem{Kostrykin:Schrader:98a} V. Kostrykin and R. Schrader, \textit{Scattering theory
approach to random Schr\"{o}dinger operators in one dimension}, Rev. Math. Phys.
\textbf{11}, 187 -- 242 (1999).

\bibitem{Kostrykin:Schrader:98b} V. Kostrykin and R. Schrader, \textit{One-dimensional
disordered systems and scattering theory}, preprint (1998); available from
\texttt{http://www-sfb288.math.tu-berlin.de/abstract/337}.

\bibitem{Kotani:Simon:87} S. Kotani and B. Simon, \textit{Localization in general one
dimensional systems. II}, Commun. Math. Phys. \textbf{112}, 103 -- 120 (1987).

\bibitem{Krengel} U. Krengel, \textit{Ergodic Theorems}, Berlin, de Gruyter, 1985.

\bibitem{Lifshitz:Gredeskul:Pastur:82} I.M. Lifshitz, S.A. Gredeskul, and
L.A. Pastur, \textit{Theory of the passage of particles and waves through
randomly inhomogeneous media}, Sov. Phys. JETP \textbf{56}, 1370 -- 1378
(1982).

\bibitem{Lifshitz:Gredeskul:Pastur:88} I.M. Lifshitz, S.A. Gredeskul, and L.A. Pastur,
\textit{Introduction to the Theory of Disordered Systems}, Wiley, New York,
1988.

\bibitem{Nakao} S. Nakao, \textit{On the spectral distribution of the Schr\"{o}dinger
operator with random potential}, Japan J. Math. \textbf{3}, 111 -- 139 (1977).

\bibitem{Pastur:71} L.A. Pastur, \textit{On the Schr\"{o}dinger equation with a
random potential}, Theor. Math. Phys. \textbf{6}, 299 -- 306 (1971).

\bibitem{Pastur:80} L.A. Pastur, \textit{Spectral properties of disordered systems
in one-body approximation}, Commun. Math. Phys. \textbf{75}, 179 -- 196 (1980).

\bibitem{Pastur:Figotin}  L. Pastur and A. Figotin, \textit{Spectra
of Random and Almost-Periodic Operators}, Berlin, Springer, 1992.

\bibitem{Pavlov:Smirnov} B.S. Pavlov and N.V. Smirnov, \textit{Spectral properties
of one-dimensional disperse crystals}, J. Sov. Math.
\textbf{31}, 3388 -- 3398 (1985).

\bibitem{RS1} M. Reed and B. Simon, \textit{Methods of Modern Mathematical Physics, I:
Functional Analysis}, New York, Academic Press, 1972.

\bibitem{RS3} M. Reed and B. Simon, \textit{Methods of Modern Mathematical Physics, III:
Scattering Theory}, New York, Academic Press, 1979.

\bibitem{RS4} M. Reed and B. Simon, \textit{Methods of Modern Mathematical Physics, IV:
Analysis of Operators}, New York, Academic Press, 1978.

\bibitem{Rorres} C. Rorres, \textit{Transmission coefficients and eigenvalues
of a finite one-dimensional crystal}, SIAM J. Appl. Math. \textbf{27}, 303
-- 321 (1974).

\bibitem{Simon:79a} B. Simon, \textit{Trace Ideals and Their Applications},
Cambridge University Press, New York, 1979.

\bibitem{Simon:79b} B. Simon, \textit{Functional Integration
and Quantum Physics}, Academic Press, New York, 1979

\bibitem{Simon:82} B. Simon, \textit{Schr\"{o}dinger semigroups}, Bull. Am. Math. Soc.
\textbf{7}, 447 -- 526; Erratum: \textit{ibid.} \textbf{11}, 426 (1984).

\bibitem{Stollmann:93} P. Stollmann, \textit{Trace ideals properties of perturbed Dirichlet
demigroups}, p. 153 -- 158 in M. Demuth, P. Exner, H. Neidhardt, and V.
Zagrebnov (Eds.), \textit{Mathematical Results in Quantum Mechanics}, Operator
Theory: Advances and Applications, Vol.70. Birkh\"{a}user, Basel, 1994.

\bibitem{Stollmann:94} P. Stollmann, \textit{St\"{o}rungstheorie von Dirichletformen mit
Anwendungen auf Schr\"{o}dingeroperatoren}, Ha\-bi\-li\-ta\-tions\-schrift (1994);
available from \texttt{http://www.math.uni-frankfurt.de/\~{}stollman/}.

\bibitem{Stollmann:Stolz:94} P. Stollmann and G. Stolz, \textit{Singular spectrum for
multidimensional Schr\"{o}dinger operators with potential barriers}, J. Oper.
Theory, \textbf{32}, 91 -- 109 (1994).

\end{thebibliography}
\end{document}